\documentclass[acmsmall,screen,nonacm]{acmart}

\AtBeginDocument{%
  }

\setcopyright{none}
\acmDOI{}
\acmISBN{}
\acmVolume{}
\acmNumber{}
\hypersetup{pdfpublisher={Authors},pdfpubtype={other}}
\makeatletter
\providecommand{\@currentauthors}{}
\let\@mktitle@i\@mktitle@iii
\renewcommand{\@mkauthors}{%
  \begingroup
  \hsize=\textwidth
  \global\setbox\mktitle@bx=\vbox{%
    \noindent\unvbox\mktitle@bx\par
    \preprintauthorblock
    \par\medskip}%
  \endgroup
}
\makeatother

\newcommand{\papertitle}{KernelAudit: Contract-Aware Auditing of Validators for LLM-Generated GPU Kernels}
\newcommand{\papershorttitle}{KernelAudit}
\newcommand{\toolname}{KernelAudit}
\newcommand{\projecturl}{https://github.com/ZIYUANLI-star/KernelAudit}

\usepackage{algorithm}
\usepackage{algpseudocode}
\usepackage{array}
\usepackage{tikz}
\usetikzlibrary{arrows.meta, positioning, fit, calc, backgrounds}

\makeatletter
\newcount\BA@floattype
\def\BA@checktopfigure#1{%
  \BA@floattype=\count#1\relax
  \divide\BA@floattype by 32\relax
  \ifnum\BA@floattype=\ftype@figure\relax
    \def\@textbottom{\vskip\z@\@plus.0001fil}%
  \fi
}
\AddToHook{cmd/@makecol/before}{%
  \let\BA@saved@textbottom\@textbottom
  \let\BA@saved@elt\@elt
  \let\@elt\BA@checktopfigure
  \@toplist
  \let\@elt\BA@saved@elt
}
\AddToHook{cmd/@makecol/after}{%
  \let\@textbottom\BA@saved@textbottom
}
\makeatother

\algrenewcommand\algorithmicrequire{\textbf{Input:}}
\algrenewcommand\algorithmicensure{\textbf{Output:}}

\newcommand{\systemname}{\textsc{\toolname}}
\newcommand{\faany}{FA$_{\mathit{any}}$}
\newcommand{\fasem}{FA$_{\mathit{sem}}$}
\newcommand{\faimp}{FA$_{\mathit{imp}}$}

\newcommand{\czero}{C0}
\newcommand{\cone}{C1}
\newcommand{\ctwo}{C2}
\newcommand{\cthreea}{C3a}
\newcommand{\cthreeb}{C3b}
\newcommand{\cthreec}{C3c}

\newcommand{\inconcl}{INCONCL}

\newenvironment{rqanswer}
  {\par\smallskip\noindent\small\ignorespaces}
  {\par\smallskip}

\begin{document}

\title[\papershorttitle]{\papertitle}

\author{Ziyuan Li}
\authornote{Equal contribution.}
\affiliation{%
  \department{School of Computer Science and Engineering}
  \institution{Southeast University}
  \city{Nanjing}
  \country{China}}
\affiliation{%
  \institution{Beijing Tongming Lake Information Technology Application Innovation Center}
  \city{Beijing}
  \country{China}}
\email{lziyuan32@gmail.com}

\author{Zihe Song}
\authornotemark[1]
\affiliation{%
  \institution{The University of Texas at Dallas}
  \city{Richardson}
  \state{TX}
  \country{USA}}
\email{Zihe.song@utdallas.edu}

\author{Xiaofang Qi}
\affiliation{%
  \department{School of Computer Science and Engineering}
  \institution{Southeast University}
  \city{Nanjing}
  \country{China}}
\email{xfqi@seu.edu.cn}

\author{Wei Yang}
\authornote{Corresponding authors: Tao Xie and Wei Yang.}
\affiliation{%
  \department{College of Computer Science and Artificial Intelligence}
  \institution{Fudan University}
  \city{Shanghai}
  \country{China}}
\affiliation{%
  \institution{Beijing Tongming Lake Information Technology Application Innovation Center}
  \city{Beijing}
  \country{China}}
\affiliation{%
  \department{Institute of Systems for Advanced Computing}
  \institution{Fudan University}
  \city{Shanghai}
  \country{China}}
\email{yang_wei@fudan.edu.cn}

\author{Tao Xie}
\authornotemark[2]
\affiliation{%
  \department{School of Computer Science and Engineering}
  \institution{Southeast University}
  \city{Nanjing}
  \country{China}}
\affiliation{%
  \institution{Beijing Tongming Lake Information Technology Application Innovation Center}
  \city{Beijing}
  \country{China}}
\affiliation{%
  \department{School of Computer Science}
  \institution{Peking University}
  \city{Beijing}
  \country{China}}
\affiliation{%
  \department{College of Computer Science and Artificial Intelligence}
  \institution{Fudan University}
  \city{Shanghai}
  \country{China}}
\affiliation{%
  \department{Institute of Systems for Advanced Computing}
  \institution{Fudan University}
  \city{Shanghai}
  \country{China}}
\email{taoxie@pku.edu.cn}

\renewcommand{\shortauthors}{Ziyuan Li et al.}
\authorsaddresses{}

\newcommand{\preprintauthorblock}{%
  \begingroup
  \centering
  \setlength{\parindent}{0pt}%
  {\normalfont\fontsize{10.5}{13}\selectfont
    \mbox{Ziyuan Li\textsuperscript{1,2,*}}\quad
    \mbox{Zihe Song\textsuperscript{3,*}}\quad
    \mbox{Xiaofang Qi\textsuperscript{1}}\quad
    \mbox{Wei Yang\textsuperscript{4,2,5,\textdagger}}\quad
    \mbox{Tao Xie\textsuperscript{1,2,6,4,5,\textdagger}}\par}
  \smallskip
  {\normalfont\small
    \textsuperscript{1}School of Computer Science and Engineering,\\
    Southeast University, Nanjing, China\par
    \smallskip
    \textsuperscript{2}Beijing Tongming Lake Information Technology Application\\
    Innovation Center, Beijing, China\par
    \smallskip
    \textsuperscript{3}The University of Texas at Dallas, Richardson, TX, USA\par
    \smallskip
    \textsuperscript{4}College of Computer Science and Artificial Intelligence,\\
    Fudan University, Shanghai, China\par
    \smallskip
    \textsuperscript{5}Institute of Systems for Advanced Computing,\\
    Fudan University, Shanghai, China\par
    \smallskip
    \textsuperscript{6}School of Computer Science, Peking University, Beijing, China\par}
  \smallskip
  {\normalfont\small
    \href{mailto:lziyuan32@gmail.com}{\nolinkurl{lziyuan32@gmail.com}},\quad
    \href{mailto:Zihe.song@utdallas.edu}{\nolinkurl{Zihe.song@utdallas.edu}},\quad
    \href{mailto:xfqi@seu.edu.cn}{\nolinkurl{xfqi@seu.edu.cn}}\\
    \href{mailto:yang_wei@fudan.edu.cn}{\nolinkurl{yang_wei@fudan.edu.cn}},\quad
    \href{mailto:taoxie@pku.edu.cn}{\nolinkurl{taoxie@pku.edu.cn}}\par}
  \endgroup
}

\begin{abstract}
The effectiveness of generating GPU kernels driven by a large language model (LLM) is highly dependent on
the quality of the used validator (used to accept or reject a generated kernel candidate against the contracts of
the given kernel-generation task), whose validation results inform reward assignment, candidate selection,
and speedup claims during kernel generation. However, an existing validator suffers from various limitations,
e.g., wrongly accepting a candidate due to missing its forbidden library use or wrongly rejecting a candidate
due to determining its permitted library use as forbidden (constraints on library uses are stated in the task
contracts).

To assess the quality of a validator against the given task contracts, in this paper, we propose an independent-audit
approach for the validator by (1) construction of audit oracle: labeling a candidate as PASS or FAIL based
on three separate dimensions in our proposed contract model (i.e., semantic correctness, implementation
integrity, and measurement integrity) and (2) auditing of the validator against the audit oracle: comparing
the labels of the candidate against the validation results produced by the validator. For semantic correctness,
reference-calibrated stress tests check output correctness on valid inputs for a candidate. For implementation
integrity, reference and candidate execution traces are used to determine specific library uses. For measurement
integrity, timing checks assess whether the reported runtime reflects the full time needed to complete the
candidate's computation. We apply our approach to assess the quality of six validators and 716 candidates
from seven corpora, including constructed controls. The evaluation results show that our approach helps
assess the existing six validators to have nontrivial false-reject rates. There is high-degree agreement between
the candidate labels produced by our audit oracle and human judgments.

\end{abstract}



\maketitle

\section{Introduction}

The effectiveness of GPU-kernel generation driven by a large language
model (LLM) depends on the quality of its
validator~\cite{ouyang2025kernelbench,li2025tritonbench,lange2025towards,liu2026dr,dai2026cuda,cheng2026musacoder}.
A validator checks generated candidates against the given task contracts;
its results guide subsequent generation, candidate selection, and speedup claims.
For example, for a task requiring a custom matrix-multiplication kernel,
the validator must reject a candidate that has library use (i.e., calling a library routine) of matrix multiplication while allowing permitted library use
for memory allocation.
Insufficient validation can reward defective candidates~\cite{skalse2022defining},
accept candidates that violate the task contracts, and support misleading
performance claims.
Existing work improves validators through numerical testing and
execution monitoring.
Numerical testing~\cite{lange2025towards} examines outputs across more
inputs and numerical conditions, while fallback checks~\cite{cheng2026musacoder}
screen for prohibited library use.
Runtime attribution, such as Atrex~\cite{yang2026llm}, measures the share
of GPU execution time spent in a candidate.
Timing defenses~\cite{deepreinforce} address unreliable performance measurements.

However, the existing work~\cite{lange2025towards,cheng2026musacoder,yang2026llm,deepreinforce} does not conduct high-quality check on a generated candidate against
the given task contracts.
Consider two kernel-generation tasks with different contracts.
An elementwise-addition task requires a custom implementation of addition
and forbids delegating that computation to a library routine.
A matrix-multiplication task requires a custom implementation of matrix
multiplication but explicitly permits a library addition routine for a
bias-addition epilogue.
Treating both uses of the same addition routine as forbidden can wrongly
reject a compliant candidate for the second task; treating both as
permitted can wrongly accept a candidate that delegates the first task's
required custom computation to the library.
Distinguishing these library uses of the same library routine requires evidence of the library call's role and
contribution to the candidate's output, interpreted under the task contracts.

\begin{figure}[t]
	\centering
	\includegraphics[width=\linewidth]{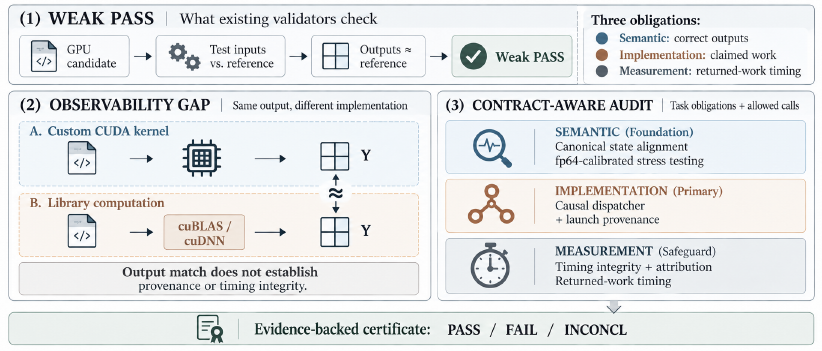}
	\caption{Three dimensions of validator acceptance.
		The audit separately checks output agreement, execution provenance,
		and timing integrity under the task contracts, preserving unresolved
		outcomes when evidence is insufficient.}
	\Description{A top row shows a GPU candidate checked against reference outputs,
		ending in a weak PASS.
		Below, two execution paths use a custom GPU kernel or a vendor library
		while producing matching tensor outputs.
		Alongside them, semantic, implementation, and measurement evidence is
		collected under the task's obligations and allowed calls.
		The audit combines the evidence into PASS, FAIL, or INCONCL certificates.
		Library use is judged under the task contract.}
	\label{fig:overview}
\end{figure}

To address the issue, we propose an independent-audit approach with two steps:
(1) constructing an audit oracle~\cite{barr2014oracle} that labels candidates
(with the PASS or FAIL certificate) along three dimensions of our contract model---semantic correctness,
implementation integrity, and measurement integrity
(Figure~\ref{fig:overview})---and
(2) comparing these labels with the validator's results.
The task contracts provide the compliance criteria, while runtime
observations and interventions provide the evidence.
For semantic correctness, reference-calibrated stress tests check output correctness on valid inputs
for a candidate. For implementation integrity, reference and candidate execution traces are used
to determine specific library uses. For measurement integrity, timing checks assess whether the
reported runtime reflects the full time needed to complete the candidate's computation. Note that
missing required evidence leaves affected checks unresolved without overriding independently
supported FAIL judgments. A PASS certificate means no detected violation within the audited scope
after all required checks and certification conditions are satisfied.

We apply our approach to assess the quality of six validators and 716 candidates from seven
corpora, including constructed controls. Among the candidates accepted by each validator and
labeled with PASS or FAIL by our audit oracle, 51.7\%--66.9\% receive FAIL. Among the 172 candidates
labeled PASS by our audit oracle, each validator rejects 9--21, yielding false-reject rates of 5.2\%--12.2\%.
An independent blind review of a stratified random sample of 120 candidates shows high-degree
agreement between the candidate labels produced by our audit oracle and human judgments. We also assess candidate selection on a fixed pool of 232 tasks.
Reusing complete primary-audit evidence for 204 slots and retaining
historical evidence for the other slots, the reconciled audit gate covers
62 tasks (26.7\%). Independent shipment labels identify 60 clean candidates,
one unsafe, and one unreviewed, bounding its unsafe fraction at 1.6\%--3.2\%.
The six existing validators cover 40.1\%--90.5\% of tasks, with
15.1\%--40.5\% of selected candidates labeled unsafe. Lower unsafe delivery
thus accompanies substantial abstention; the evidence refresh is not a new
GPU run or a prospective blind study.

The paper makes the following main contributions:
\begingroup
\setlength{\listisep}{3pt}
\begin{itemize}
	\item A contract model separating semantic correctness, implementation
	integrity, and measurement integrity.
	\item An independent audit combining execution evidence with task
	contracts and preserving unresolved outcomes.
	\item A systematic assessment of six validators and audit
	judgments against independent human review.
	\item An evaluation of unsafe delivery, coverage, and candidate
	rankings under audit-based selection.
\end{itemize}
\endgroup

We release the code and replication materials for \systemname{} at
\expandafter\url\expandafter{\projecturl}.

\section{Background, Contracts, and Research Questions}
\label{sec:contracts-taxonomy}

\paragraph{Task Contracts.}
A task contract specifies the semantic, implementation, and measurement obligations for a candidate.
The contract defines valid inputs, a reference, permitted library \emph{mediators} (routines computing on the candidate's behalf), and a timing policy.
For candidate $k$ on task $t$, $S_t(k)$ requires reference-matching outputs on valid inputs under a numerically justified tolerance.
$I_t(k)$ forbids undeclared framework or vendor-library mediation of core work when a custom kernel is required.
$M_t(k)$ requires latency to cover the work producing the returned output.
Compliance requires all three obligations: $\mathit{clean}_t(k) \equiv S_t(k) \land I_t(k) \land M_t(k)$.

Task rules specify permitted library uses.
An undeclared \texttt{aten::mm} performing a custom GEMM's dominant computation violates the implementation contract even when outputs match.
Contract-allowed mediators and epilogue glue remain exempt (Section~\ref{subsec:impl-tier}).

\paragraph{Audit Outcomes.}
Audit outcomes relate certificates to a weak verifier's decision $W$ (Table~\ref{tab:outcome-taxonomy}).
A PASS certificate records no detected violation on audited surfaces, without establishing $\mathit{clean}_t(k)$.

\begin{table}[tb]
\caption{Outcome taxonomy.
\czero{} is weak-agnostic.
\cone{} pairs a weak rejection with a FAIL certificate.
\ctwo{}--\cthreec{} are verifier-relative false-accept labels assigned only when $W$ accepts.}
\label{tab:outcome-taxonomy}
\footnotesize
\setlength{\tabcolsep}{3pt}
\begin{tabular}{@{}p{0.26\linewidth}p{0.70\linewidth}@{}}
\toprule
Class & Operational condition \\
\midrule
\czero{} Certified & Oracle certificate PASS
(Section~\ref{subsec:conservative-aggregation}): no violation detected on
any required axis; in false-accept tables, also $W(t,k)=\mathrm{PASS}$. \\
\cone{} Public failure & $W$ rejects \emph{and} the certificate is FAIL
(weak rejections of PASS certificates are false rejects, reported
separately). \\
\ctwo{} Unattributed silent defect & $W$ accepts, $S_t(k)$ fails, no
qualifying mechanism signature. \\
\cthreea{} Output-overfitting violation & $W$ accepts, $S_t(k)$ fails, and a
hardcode/tolerance/specialization signature maps to $W$'s blind spot. \\
\cthreeb{} Implementation violation & $W$ accepts and $I_t(k)$ fails:
dispatcher-confirmed output dependence on policy-forbidden work or a
non-mediated vendor launch. \\
\cthreec{} Measurement violation & $W$ accepts and $M_t(k)$ fails on a
decisive, positively controlled guard (Section~\ref{subsec:meas-tier}). \\
\inconcl{} Abstention & An auditability precondition remains unmet; no
evidence to convict or acquit. Carries a root cause. \\
\bottomrule
\end{tabular}
\end{table}

Outcome codes distinguish violations by their supporting evidence.
C3 requires an axis FAIL, an executable signature, and a mapping to $W$'s blind spot.
Other semantic-only failures remain \ctwo{}.
Because axis evidence can overlap, certificate-level false accepts and RQ4 slot ledgers use the priority \cthreeb{} over \cthreec{} over \cthreea{} over \ctwo{}.
The priority is a reporting convention, not an ordering of severity or causal strength.

\paragraph{Research Questions.}
Four research questions examine validator acceptance, audit validity, performance claims, and candidate selection.
\textbf{RQ1 (Prevalence and structure):} among oracle-decided records, what fraction of accepted records has a FAIL certificate, what fraction of FAIL certificates does a validator accept, and are the main misses parametric or structural?
\textbf{RQ2 (Attribution and oracle validity):} what axes and channels support FAIL decisions, and how do audit labels compare with the blind reference in sensitivity, specificity, coverage, and abstention?
\textbf{RQ3 (Credibility and certification):} how do reportable local speedups compare with upstream claims, and what fraction of the paired records receives a PASS certificate?
\textbf{RQ4 (Training-free remedy):} how do gate and ordering changes affect unsafe delivery, coverage, speed, and ranking on the fixed pool, and what audit-cost evidence is available?

\section{A Conservative, Evidence-Tiered Audit Oracle}
\label{sec:method}

\systemname{} implements the audit oracle, which audits a candidate against its reference and task contract, returning three axis verdicts and a certificate.
The audit combines reference-based testing, runtime observation, and program intervention on contract-supported inputs.
Task rules determine whether the evidence supports conviction, withholds certification, or remains a non-gating signal.

\subsection{Verdict Semantics}
\label{subsec:verdict-semantics}

All tasks require the three axes, collected in $R_t$.
PASS means no detected violation on audited surfaces; FAIL requires
candidate-attributable violation evidence. INCONCLUSIVE denotes unmet
auditability preconditions; UNSUPPORTED denotes unavailable capabilities.
Neither satisfies a required axis. Missing preconditions limit their channel
without erasing an independent FAIL; tool failures alone cannot convict
(Table~\ref{tab:preconditions}).

The main audit runner gives each candidate a fresh seeded worker;
candidate-specific measurement conviction requires isolated attribution (R-C).
RQ4 uses a shared-process runner and separately defined replay rules for
saved verdicts, eligibility, and timing (Section~\ref{sec:selector-intervention}).
Guards protect shared canonical inputs, module state, and execution policy;
candidates may mutate their private input clones.

Thresholds and the vendor inventory are frozen and digest-checked, with fail-closed loading.
Appendices~\ref{s:params}--\ref{s:annotation} provide the parameter and protocol details.

\begin{table}[tb]
\caption{Auditability conditions and affected decisions. An unresolved channel
does not erase an independent FAIL. Certificate and speedup eligibility are
separate; axis-level evidence categories appear in
Section~\ref{subsec:inconclusive} and Appendix~\ref{s:inconcl}.}
\label{tab:preconditions}
\footnotesize
\setlength{\tabcolsep}{3pt}
\begin{tabular}{@{}p{0.68\linewidth}p{0.28\linewidth}@{}}
\toprule
Unmet condition or trigger & Effect (scope) \\
\midrule
Build/import failure or resource exhaustion leaves no executable audit path &
\inconcl{} (execution) \\
No reliable state mapping & \inconcl{} (semantic) \\
Calibrated allowance $\bar\rho_\ell>\rho_{\max}$ & \inconcl{} (semantic) \\
Crash or nondeterminism prevents a comparable replay & \inconcl{} (affected channel) \\
\textbf{R-A}: no positive work evidence & \inconcl{} (implementation) \\
\textbf{R-B}: listed degenerate reference & certificate withheld (task) \\
Localization replay missing or incomparable & \inconcl{} (causal probe) \\
\textbf{R-C}: unattributed decisive timing signal in a shared process &
\inconcl{} (measurement) \\
Stream differential meets its criterion & certificate withheld \\
A/A self-comparison unstable & speedup withheld \\
\bottomrule
\end{tabular}
\end{table}

\subsection{Foundation Tier: Semantic Axis}
\label{subsec:semantic-axis}

Semantic comparison requires verified candidate--reference state correspondence:
same-seed instantiation need not ensure equal RNG consumption or weights in
restructured models. We inject canonical reference weights using a mapping verified for keys,
shapes, dtypes, values, and coverage. Unresolved correspondence or undeclared
candidate-only state causes abstention. Valid inputs and a usable reference remain
necessary after alignment (Appendix~\ref{s:inconcl}).

Empirical studies document value- and dtype-sensitive numerical failures in
GPU programs~\cite{rathnasuriya2025gpunumeric} and shape-, dtype-, and
backend-sensitive code-generation bugs in tile-programming
frameworks~\cite{rathnasuriya2026tilebugs}.
Motivated by high-precision reference checking~\cite{sarkar2026correctness}, we calibrate each input's floating-output tolerances using state-identical native-precision and fp64 reference executions.
In fp32, let $d_{\ell i}=|r^{32}_{\ell i}-r^{64}_{\ell i}|$.
For each floating leaf $\ell$, $\epsilon^{\mathrm{abs}}_\ell$ and
$\epsilon^{\mathrm{rel}}_\ell$ are the 99th percentiles of $d_{\ell i}$ and
$d_{\ell i}/\max(|r^{64}_{\ell i}|,10^{-12})$, respectively, over that leaf's
elements, without pooling across leaves, inputs, or runs.
With the operator-family multiplier fixed at one, set
$\bar\rho_\ell=c\epsilon^{\mathrm{rel}}_\ell$,
$\rho_\ell=\max(\rho_{\min},\bar\rho_\ell)$, and
$a_\ell=\max(a_{\min},c\epsilon^{\mathrm{abs}}_\ell)$.
For candidate output $k$ and native-precision reference $r$ ($r=r^{32}$ here),
\begin{equation}
\operatorname{NumPass}_\ell(k,r)\Longleftrightarrow
\forall i:\ |k_{\ell i}-r_{\ell i}|\leq a_\ell+\rho_\ell|r_{\ell i}|.
\label{eq:semantic-rule}
\end{equation}
Here $c=4$, $\rho_{\min}=10^{-5}$, $\rho_{\max}=10^{-3}$, and
$a_{\min}=10^{-8}$. The comparison remains unresolved if any
$\bar\rho_\ell>\rho_{\max}$, the required reference precision is unsupported,
or the reference output is nonfinite.
Otherwise, the rule applies per leaf with exact structure and dtype checks.
The multiplier expands the reference-discrepancy scale, while the floors retain
a minimum allowance. The ceiling makes excessive relative calibration an
auditability limit rather than clipping it into an accepted tolerance.
These settings define an empirical acceptance policy, not a statistical error
bound; fp64 is a calibration reference rather than exact truth.

Our 16-slot schedule draws on prior stress testing~\cite{lange2025towards} to cover contract-supported value, distribution, shape/dtype, layout, repeated-call, and provider-seed tests.
Unsupported optional transformations use native-seed tests
(Appendix~\ref{s:miss-matrix}).
A candidate NaN or infinity against a finite reference is a semantic failure;
an incomparable execution leaves the affected evidence unresolved.
Coverage remains limited to the exercised inputs and call sequences: the
stale-cache and specialization controls illustrate unexercised behavior
(Table~\ref{tab:controls}).

\subsection{Primary Tier: Implementation Axis}
\label{subsec:impl-tier}

\begin{figure}[!t]
\centering
\includegraphics[width=\linewidth]{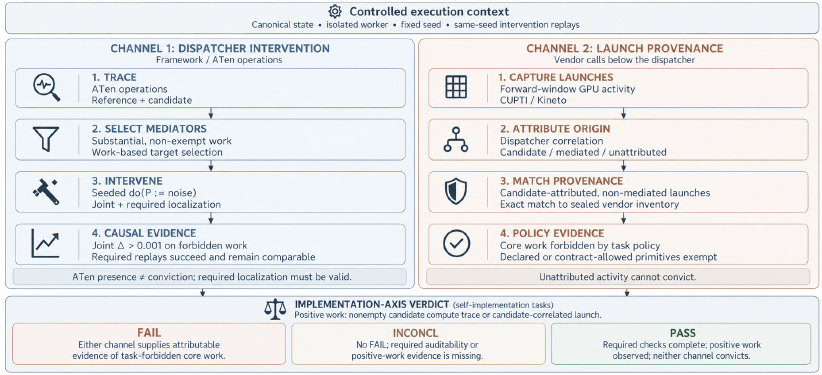}
\caption{Complementary implementation evidence under a fixed seed. Dispatcher
interventions test output dependence; launch provenance examines vendor activity
beyond observed ATen mediation. Conviction requires attributable evidence of
work forbidden by the task contract.}
\Description{Two parallel evidence channels within a controlled execution
context. The dispatcher channel selects suspect operations and compares clean,
joint, and localized replays. The launch channel attributes observed forward-window
activity and checks non-mediated vendor launches against a sealed inventory.
The implementation decision gives FAIL priority when either channel supplies
attributable evidence of contract-forbidden core work. Without FAIL, it abstains
if required auditability or positive-work evidence is missing; PASS requires
complete checks, observed candidate work, and no conviction.}
\label{fig:impl-axis}
\end{figure}

Matching outputs cannot reveal delegation. Two channels observe ATen-mediated
computation and vendor calls below the dispatcher
(Figure~\ref{fig:impl-axis}). Static source tokens guide probes but cannot convict.

\paragraph{Channel 1: Dispatcher Causal Probe.}
Dispatcher traces select suspect mediators by compute-class work, measured with FLOP estimates or element-count proxies rather than GPU time.
Let $R$ be total reference work, $C_g$ candidate ATen work in class $g$ after excluding identifiable contract-allowed calls, and $C=\sum_g C_g$.
The oracle selects all non-exempt operations in a class that matches the reference's dominant class and reaches 5\% of that class's work.
A class also qualifies when $C_g\geq0.30R$, or when both $C\geq0.05\max(R,1)$ and $C_g\geq0.30C$.
These rules select probes, while conviction also requires output dependence on forbidden work.
The unit floor provides numerical protection.
Coverage depends on work-proxy accuracy and operation support (Appendix~\ref{s:cases}).

Comparable program interventions use causal reasoning~\cite{pearl2003causality} and testing~\cite{johnson2020causal} to examine whether returned values depend on selected mediators.
\texttt{do(P := noise)} denotes replacing targeted primary floating outputs $P$
in place with statistically matched, deterministically seeded noise.
Let $y$ be the clean output and
$\tilde y$ the output under the joint intervention on all targets:
\begin{equation}
\Delta(y,\tilde y)=\max_{\ell,i}
\frac{|y_{\ell i}-\tilde y_{\ell i}|}{\max(|y_{\ell i}|,10^{-12})}.
\label{eq:causal-score}
\end{equation}
This relative score measures output change, not significance or delegated work;
near-zero clean outputs can amplify small absolute changes.

Channel 1 records FAIL when joint output change exceeds $10^{-3}$ on
contract-forbidden work. Required replays use aligned modules and inputs,
must execute successfully, perturb every target selected for that replay, and preserve output
structure, shape, and dtype. Violations of protected execution state also
invalidate the probe. With multiple targets and a joint effect, every required
localization must pass these checks; otherwise the probe is INCONCLUSIVE.
Joint and individual perturbations test dependence on the selected set and
effects within it, respectively. A joint effect need not appear individually
and does not identify a unique mechanism.
The replay checks establish observable comparability, not control-flow
equivalence (Appendix~\ref{s:params}).

\paragraph{Channel 2: Kernel-Launch Provenance.}
To examine delegation bypassing the dispatcher, the oracle collects
forward-window GPU launches through
CUPTI/Kineto~\cite{nvidia2025profiling} and aligns them using dispatcher correlation.
ATen-mediated launches enter the mediated-library bucket for Channel~1;
their presence alone cannot convict. Remaining launches require candidate
attribution; unattributed activity is retained without conviction.

Candidate-attributed launches without an ATen mediator are matched exactly
against 23{,}666 digest-pinned cuBLAS and cuDNN symbols. Attribution, a vendor
match, and the frozen policy jointly support FAIL when the observed work is
forbidden core computation. The match supplies occurrence evidence; the policy
determines whether that work is prohibited. Declared or contract-allowed
primitives remain exempt.
Coverage is bounded by the sealed inventory (Section~\ref{sec:threats}).

\paragraph{Work Presence and Executed Fallbacks.}
Implementation PASS requires no detected forbidden work plus positive work
evidence: a nonempty candidate compute trace or candidate-correlated launch (R-A).
Work presence alone does not establish compliance.

R-B withholds certification for listed degenerate-reference tasks:
matching outputs lack the informativeness this policy requires
(Section~\ref{subsec:policy-v31}).

After a custom-build failure, both channels audit the executed native-PyTorch
fallback. Attributable forbidden delegation supports FAIL; finding no violation
on that path cannot certify the unavailable custom implementation
(Section~\ref{subsec:inconclusive}).

\subsection{Safeguard Tier: Measurement Axis}
\label{subsec:meas-tier}

For the published timing-threat catalogue~\cite{deepreinforce}, validation and attribution evidence distinguish conviction channels, certification-withholding guards, and non-gating signals.
Thread observation, lazy outputs, and static low-precision casts remain non-gating signals.

Before candidate import, timing-integrity checks snapshot the identities of \texttt{time.perf\_counter}, \texttt{torch.cuda.Event} methods, and \texttt{synchronize}.
Positively controlled checks support measurement FAIL for attributable changes under isolated execution and controlled harness state (Table~\ref{tab:controls}).

The stream differential can withhold certification without establishing a
measurement violation. Randomized ABBA comparisons use timing with and without
the synchronization guard. Let $\widehat R$ denote the overall guarded median
divided by the overall unguarded median. Withholding requires $\widehat R\geq3$, a paired-batch
95\% percentile bootstrap interval with lower endpoint at least $3$,
an absolute median gap $\geq0.05$ ms, and attributable evidence
(Appendix~\ref{s:params}). This channel never convicts because its positive control
does not demonstrate convicting power (Section~\ref{subsec:measurement-axis-results}).

A/A screens timing instability~\cite{mytkowicz2009producing} with two reference
instances: 10 warmups and three paired 20-timing batches per arm in
ABBA/BAAB/ABBA order. Both CVs over all 60 latencies and the maximum normalized
within-batch order effect must be $\leq0.05$. A paired bootstrap of the three
batch-median pairs gives a nominal 95\% percentile ratio interval $[L,U]$;
passage requires $L\leq1\leq U$, $U-L\leq0.1$, and internal timing PASS,
including output/state consistency. Appendix~\ref{s:params} defines the odd/even
and half-batch order contrasts and resampling. This screen withholds speedups,
not certificates, and does not guarantee unbiased timing.

\subsection{Aggregation and Certification}
\label{subsec:conservative-aggregation}

Within one complete audit record, positive violation evidence takes precedence.
Let $V_a(k;e)$ be the verdict on required axis $a\in R_t$ in record $e$,
where $R_t\neq\varnothing$. Write
$\mathcal U=\{\mathrm{INCONCLUSIVE},\allowbreak\mathrm{UNSUPPORTED}\}$ for the unresolved
statuses. The axis aggregate is
\begin{equation}
A_t(k;e)=
\begin{cases}
\mathrm{FAIL}, & \exists a\in R_t:\ V_a(k;e)=\mathrm{FAIL},\\
\mathrm{INCONCL}, & \nexists a\in R_t:\ V_a(k;e)=\mathrm{FAIL}
\ \land\ \exists a\in R_t:\ V_a(k;e)\in\mathcal U,\\
\mathrm{PASS}, & \forall a\in R_t:\ V_a(k;e)=\mathrm{PASS}.
\end{cases}
\label{eq:certificate}
\end{equation}

R-B and the stream guard can withhold certification even when all required axes pass.
Let $Q_t(k;e)$ denote eligibility under these guards.
The certificate is $C_t(k;e)=A_t(k;e)$, except that $A_t(k;e)=\mathrm{PASS}$ and $Q_t(k;e)$ false yield $C_t(k;e)=\mathrm{INCONCL}$.
Certificate withholding leaves axis verdicts unchanged.
R-A/R-C affect axis verdicts.
The A/A gate affects speedups only.

The release manifest binds each certificate to one complete record after eligibility checks.
Diagnostic axes may cite separate qualified executions but cannot substitute for the certificate-bound evidence.
Appendix~\ref{s:params} specifies source qualification, export, and restricted-merge rules.
Selection predicates read neither weak decisions nor human-label values.
Certificates determine counts, while selected axes supply diagnostics.

\section{Experimental Design}
\label{sec:experimental-design}

The experiments use 716 task--candidate--reference records for RQ1--RQ3 and a separate fixed candidate pool for RQ4.
The audit records span seven corpora (Table~\ref{tab:contract-provenance}).
The two TritonBench-G reference layers reuse source kernels.
The candidate-selection pool contains 232 tasks and 1{,}160 candidate slots.

\begin{table}[tb]
\caption{Audited corpora, including records with \inconcl{} certificates.}
\label{tab:contract-provenance}
\footnotesize
\setlength{\tabcolsep}{3pt}
\begin{tabular}{@{}p{0.30\linewidth}r p{0.60\linewidth}@{}}
\toprule
Corpus & N & Provenance and role \\
\midrule
CUDA-L1~\cite{sun2026cuda} & 241 & Main generated-CUDA corpus; KernelBench-style
references, provider inputs, custom-replacement policy. \\
AI-CUDA-Engineer~\cite{lange2025ai} & 229 & Main generated-CUDA corpus;
released task sources and references, same policy. \\
KernelBenchX~\cite{wang2026kernelbenchx} & 63 & External generated-Triton corpus
with upstream-independent references. \\
TritonBench-G~\cite{li2025tritonbench}, self-built ref.\ layer & 89 & Production
Triton kernels; our own independent algorithmic references,
cross-validated on CPU. \\
TritonBench-G, upstream ref.\ layer & 26 & Same source kernels with the
upstream repository's own references (reference-quality probe). \\
cuDNN/cuBLAS delegation controls & 60 & Positive implementation controls:
candidates explicitly call cuDNN (35) / cuBLAS (25) behind independent
references. \\
robust-kbench handwritten~\cite{lange2025towards} & 8 & Curated handwritten
set, upstream-presumed clean (probed in Section~\ref{subsec:impl-channels}). \\
\bottomrule
\end{tabular}
\end{table}

\paragraph{Subjects and Task Policies.}
A frozen \emph{contract manifest} specifies implementation policies, provenance,
and permitted non-core glue (policy resolution in Appendix~\ref{s:params}).
The main CUDA corpora require core self-implementation under our frozen policy;
their upstream prompts establish no blanket ban on core library use. Permitted glue includes allocation,
initialization, dtype/shape handling, and small epilogues. Constructed delegation
controls explicitly require self-implementation. For TritonBench-G and
robust-kbench, this requirement is our strict reading of
``self-contained/\allowbreak handwritten,'' recorded as \texttt{author-strict-reading}.
Implementation labels follow these policies and
Section~\ref{subsec:impl-tier}'s observation rules;
Section~\ref{subsec:oracle-validity} distinguishes the human rubric.

\paragraph{Evaluated Verifiers.}
Of the six weak verifiers (Table~\ref{tab:weak-verifier-provenance}),
CUDA-Agent, KernelBench, KernelGYM,
and robust-kbench use source-pinned adapters; DeepReinforce and MusaCoder
use \emph{publication-derived} reconstructions. The artifact documents versions
and fidelity. Results characterize these adapters, including omissions.

\begin{table}[tb]
\caption{Evaluated checking surfaces. Each listed tensor-comparison tolerance
$t$ sets $\mathrm{atol}=\mathrm{rtol}=t$.
The Atrex-based row is an implementation-only RQ2 comparator ($n{=}120$ records);
\emph{Not evaluated} denotes study scope, not tool capability.}
\label{tab:weak-verifier-provenance}
\footnotesize
\setlength{\tabcolsep}{3pt}
\begin{tabular}{@{}p{0.17\linewidth}p{0.33\linewidth}p{0.28\linewidth}p{0.17\linewidth}@{}}
\toprule
Method & Semantic check & Implementation check & Measurement check \\
\midrule
CUDA-Agent~\cite{dai2026cuda} & \texttt{assert\_close} $10^{-2}$, 5 trials &
runtime blocking of \texttt{F.*} & none \\
KernelBench~\cite{ouyang2025kernelbench} & \texttt{allclose} fp32 $10^{-4}$;
fp16/bf16 $10^{-2}$, 5 trials &
none & none \\
KernelGYM~\cite{liu2026dr} & \texttt{allclose} $10^{-2}$, 5 trials &
none in the evaluated adapter (upstream decoy signal not included) & none \\
robust-kbench~\cite{lange2025towards} & \texttt{allclose} $10^{-5}$, 5 trials,
seeded, double evaluation & none & none \\
DeepReinforce~\cite{deepreinforce} & \texttt{allclose} $10^{-2}$, 5 trials &
none & 4 guards (timing, materialization, threads, stream) \\
MusaCoder~\cite{cheng2026musacoder} & \texttt{allclose} $10^{-2}$, 5 trials & static
high-level fallback scan (within a five-category policy) & none \\
\midrule
Atrex-based screen~\cite{yang2026llm} & Not evaluated &
Attribution screen only: target-kernel GPU time / end-to-end latency;
$\leq 0.5$ suspicious, not an implementation-axis verdict &
Not evaluated as measurement integrity; timing is an attribution input \\
\bottomrule
\end{tabular}
\end{table}

\paragraph{Verdict Alignment.}
Binary weak decisions are compared with three-valued audit certificates on
569 \emph{oracle-decided} records (3{,}414 record--verifier pairs).
Weak REJECT includes failed checks and inability to evaluate.

\paragraph{Metrics and Evidence.}
The \emph{false-accept rate} is the fraction of weak-accepted, oracle-decided
records with a FAIL certificate. The \emph{miss rate} is the accepted
fraction of FAIL certificates; the \emph{false-reject rate} is the rejected
fraction of PASS certificates. Selected-axis FAIL rates share the false-accept
denominator but may use separate valid evidence; they diagnose individual axes
without decomposing certificate outcomes.

RQ3 assigns removed records by semantic $>$ implementation $>$ measurement priority, differing from Section~\ref{sec:contracts-taxonomy}'s outcome-code priority.

\paragraph{Study Denominators.}
\emph{Decidability} is (PASS$+$FAIL)$/N$; RQ3's \emph{measurable} records
have reportable, positive speedups. In RQ4, \emph{eligibility} divides admitted
slots by all slots; \emph{coverage} divides shipped tasks by all tasks.

\paragraph{Statistical Protocol.}
Counts describe fixed audited populations conditional on recorded decisions.
Blind-review point estimates use frozen sampling weights; 95\% intervals
quantify model-based posterior uncertainty under finite-population
Jeffreys--Dirichlet completion (Appendix~\ref{s:annotation}), sharing label probabilities
within strata. Timing stability and intervals follow
Section~\ref{subsec:meas-tier} (resampling in Appendix~\ref{s:params}).

Paired McNemar tests~\cite{mcnemar1947note} use Holm correction~\cite{holm1979simple} to adjust exact binomial $p$-values.
Verifier tests pair acceptance on FAIL-certificate records separately for the natural-five pool (first five corpus blocks in Table~\ref{tab:contract-provenance}) and all seven blocks.
Selector tests would pair unsafe-shipment outcomes by task; the refreshed
RQ4 comparisons are withheld because one selected candidate lacks a human label.
Tests assume independent paired units.
Source reuse across reference layers can violate independence, so verifier $p$-values are record-level reference calculations rather than independent-kernel inference (Section~\ref{sec:threats}).

Decided-record error rates exclude \inconcl{}. Best- and worst-case
false-accept bounds hold weak decisions and resolved certificates fixed
while assigning unresolved accepted records to PASS or FAIL (Appendix~\ref{s:miss-matrix}).
These bounds do not correct errors in resolved labels.
RQ4 retains all 232 tasks in shipping-rate and coverage denominators,
including abstentions, and separately reports unsafe shipments among actual
deliveries. RQ4 bounds assign its one unknown shipment clean or unsafe; they
are identification bounds, not confidence intervals.

\paragraph{Use of AI Tools (Disclosure).}
AI-generated and released candidates receive the same audit policy.
Claude Opus 4.5 generated 696 RQ4 candidates; prompts and settings
are in the artifact. Human annotators establish the blind reference labels.
AI coding assistants helped engineer tooling and prepare the manuscript.
OpenAI image-generation tools assisted layout and rendering of
Figures~\ref{fig:overview} and~\ref{fig:impl-axis}.

\paragraph{Infrastructure.}
Local experiments use an NVIDIA A800 GPU (80\,GB), PyTorch 2.1.0,
CUDA 12.1, and Python 3.10 on Ubuntu 22.04. Settings are fixed per
experiment. The primary audit uses fresh seeded workers; historical RQ4
evidence uses a shared-process batch runner. The RQ4 refresh reuses complete
primary-audit evidence for 204 slots without new GPU execution
and retains historical evidence for 956 slots. GPU clocks on the shared timing host are
unlocked (Appendix~\ref{s:params}).

\section{RQ1: Prevalence and Structure of False Accepts}
\label{sec:results}

\begin{table}[tb]
\caption{Oracle decisions by corpus. Selected axis-FAIL counts may overlap
and cite different executions. Certificate PASS/FAIL/\inconcl{} columns
partition each corpus using their own source records and eligibility conditions
(Section~\ref{subsec:conservative-aggregation}); aggregating the displayed
axes does not reconstruct them.}
\label{tab:oracle-landscape}
\footnotesize
\setlength{\tabcolsep}{3.5pt}
\begin{tabular}{@{}lrrrrrrr@{}}
\toprule
Corpus & N & Sem.\ FAIL & Impl.\ FAIL & Meas.\ FAIL & PASS & FAIL & \inconcl{} \\
\midrule
CUDA-L1 & 241 & 25 & 136 & 0 & 2 & 148 & 91 \\
AI-CUDA-Engineer & 229 & 35 & 93 & 0 & 94 & 120 & 15 \\
KernelBenchX & 63 & 11 & 22 & 0 & 23 & 32 & 8 \\
TritonBench-G (self-built ref.) & 89 & 14 & 7 & 0 & 36 & 21 & 32 \\
TritonBench-G (upstream ref.) & 26 & 16 & 1 & 0 & 8 & 17 & 1 \\
Delegation controls & 60 & 16 & 52 & 0 & 4 & 56 & 0 \\
robust-kbench handwritten & 8 & 1 & 2 & 0 & 5 & 3 & 0 \\
\midrule
Total & 716 & 118 & 313 & 0 & 172 & 397 & 147 \\
\bottomrule
\end{tabular}
\end{table}

\paragraph{Two Conditional Rates.}
\looseness=-1
False-accept and miss rates share a numerator but condition on accepted, oracle-decided records and FAIL certificates, respectively (Section~\ref{sec:experimental-design}).
Both use study contracts that may differ from generators' acceptance rules.

\paragraph{Decidability.}
\looseness=-1
The oracle decides 569/716 records (79.5\%; Table~\ref{tab:oracle-landscape}).
CUDA-L1 has the lowest decidability, 150/241 (62.2\%); the other corpora
range from 64.0\% to 100.0\%. The rates below describe decided subsets;
Section~\ref{subsec:inconclusive} examines the unresolved evidence.

\begin{table}[tb]
\caption{False accepts by corpus and pool. Oracle cells $a/b$ count FAIL
certificates among accepted, oracle-decided records. Pooled \faany{} and
selected-axis \fasem{}/\faimp{} share that denominator; axes do not decompose
certificates. Measurement-axis FA is zero. Human rows give design-weighted
estimates and 95\% model-based posterior intervals for the same main-corpus
decided-accepted sets under the independent rubric. Both pools exclude controls.
Certificate detail and abstention bounds: Appendix~\ref{s:miss-matrix}; per-corpus axis rates: released data.
robust-kb.\ $=$ robust-kbench, DeepReinf.\ $=$ DeepReinforce.}
\label{tab:main-fa}
\footnotesize
\setlength{\tabcolsep}{3pt}
\begin{tabular}{@{}lrrrrrr@{}}
\toprule
Corpus (decided $N$) & CUDA-Agent & KernelBench & KernelGYM & robust-kb.\ &
DeepReinf.\ & MusaCoder \\
\midrule
CUDA-L1 (150) & 23/23 & 117/117 & 136/138 & 110/110 & 75/75 & 39/39 \\
AI-CUDA-Engineer (214) & 97/191 & 86/174 & 97/191 & 83/167 & 98/192 &
98/192 \\
KernelBenchX (55) & 17/32 & 21/38 & 23/40 & 20/37 & 23/40 & 16/32 \\
TB-G self-built (57) & 8/40 & 15/49 & 15/49 & 10/44 & 14/48 & 12/46 \\
TB-G upstream (25) & 1/8 & 1/8 & 1/8 & 1/8 & 1/8 & 1/8 \\
Delegation controls (60) & 55/59 & 56/60 & 56/60 & 39/43 & 56/60 & 1/3 \\
rk handwritten (8) & 2/7 & 2/7 & 2/7 & 2/7 & 2/7 & 2/7 \\
\midrule
Main (364) & 120/214 & 203/291 & 233/329 & 193/277 & 173/267 & 137/231 \\
\quad \faany{}\,\% & \textbf{56.1} & \textbf{69.8} & \textbf{70.8} &
\textbf{69.7} & \textbf{64.8} & \textbf{59.3} \\
\quad Sem.-evidence / Impl.-evidence \% & 8.4/51.9 & 4.1/67.4 & 9.1/66.6 & 4.0/67.5 & 10.1/59.2 &
8.7/55.0 \\
\quad Human FA\,\% & 44.2 & 49.7 & 51.6 & 46.1 & 50.6 & 50.6 \\
\quad 95\% posterior interval & [29.9,50.0] & [33.3,53.3] & [37.4,56.8] &
[28.5,50.5] & [36.7,57.3] & [35.5,55.4] \\
\midrule
Natural-5 (501) & 146/294 & 240/386 & 272/426 & 224/366 & 211/363 &
166/317 \\
\quad \faany{}\,\% & 49.7 & 62.2 & 63.8 & 61.2 & 58.1 & 52.4 \\
\quad Sem.-evidence / Impl.-evidence \% & 8.8/43.2 & 6.2/57.0 & 10.3/57.0 & 5.2/57.7 & 11.0/50.1 &
10.4/44.8 \\
\bottomrule
\end{tabular}
\end{table}

\paragraph{Prevalence: False Accepts.}
\looseness=-1
Under the study contracts, pooling all seven corpora, including controls and handwritten kernels, yields false-accept rates of 51.7\%--66.9\%.
The two main generated-CUDA corpora and the natural-five pool yield 56.1\%--70.8\% and 49.7\%--63.8\%, respectively (Table~\ref{tab:main-fa}).
Pooled rates weight corpora by each verifier's decided acceptance counts.

\paragraph{Distribution of Axis Evidence.}
\looseness=-1
The balance of selected axis-FAIL evidence varies by corpus.
Implementation evidence exceeds semantic evidence on the main corpora (\faimp{} 51.9\%--67.5\% versus \fasem{} 4.0\%--10.1\%) and KernelBenchX (37.5\%--51.4\% versus 2.7\%--12.5\%).
The ordering reverses for TritonBench-G's self-built reference layer (\fasem{} 10.0\%--20.4\% versus \faimp{} 4.3\%--9.1\%).
The handwritten set also contains two accepted records with implementation-FAIL evidence under our strict contract reading (Section~\ref{subsec:impl-channels}).

\paragraph{Unresolved Records and Independent Labels.}
\looseness=-1
Sensitivity bounds extend the main-pool analysis to all accepts by assigning accepted \inconcl{} records to PASS or FAIL.
The lower endpoints range from 49.2\% to 60.3\% and the upper endpoints from 61.5\% to 75.3\%, holding resolved certificates fixed without correcting their errors.
Table~\ref{tab:main-fa} separately reports OV120's weighted human-referenced estimates (44.2\%--51.6\%) and posterior intervals for the decided-accepted sets.
Section~\ref{subsec:oracle-validity} gives the sampling, model, and rubric differences.

\paragraph{Miss Rates and False Rejects.}
\looseness=-1
On the main pool, verifiers accept 120--233 of 268 FAIL-certificate
records (miss rates 44.8\%--86.9\%) and reject 0--12 of 96 PASS records.
Natural-five miss rates are 43.2\%--80.5\% among 338
FAIL records, with 9--21 false rejects among 163 PASS records.
Under an independent-record model, 14 of 15
McNemar contrasts have Holm-adjusted $p<0.05$; the exception is
robust-kbench versus DeepReinforce. Source reuse across reference layers
limits this inference (Section~\ref{sec:experimental-design}; Appendix~\ref{s:miss-matrix}).

\paragraph{Constructed Controls and Ranking.}
Constructed controls change verifier rankings because their delegation pattern matches MusaCoder's fallback scan.
MusaCoder accepts one of 56 FAIL-certificate controls, compared with 39--56 for other verifiers.
Adding controls to the natural-five pool gives 394 FAIL records and miss rates of 42.4\%--83.2\%.
MusaCoder moves from second-lowest (166/338, 49.1\%) to lowest (167/394, 42.4\%).
The handwritten set's three FAILs are excluded from this pool but enter the all-seven-corpus analysis ($n{=}397$; Appendix~\ref{s:miss-matrix}).

\paragraph{Numerical Signals and Test Coverage.}
\label{subsec:parametric-structural}
The nominal-input diagnostic uses stored evidence for 62 semantic-miss
records, including 16 delegation controls. Across 15 policies per record,
57 records have at least one maximum relative error above $10^{-5}$,
four further records have non-numeric signals, and one has neither
(Appendix~\ref{s:miss-matrix}).
The same relative-error criterion flags 69/172 PASS-certificate records
(40.1\%). Because it omits the absolute-tolerance term, these flags are not
measured rejections by the full comparator.

The remaining record, \texttt{exp\_sqrt}, passes the first repeated-run
invocation but fails the second and third (maximum relative error 17.8).
The nominal diagnostics miss this repeated-execution failure.

\begin{rqanswer}\textbf{Answer to RQ1.}
Across six verifiers, 56.1\%--70.8\% of accepted, oracle-decided main-corpus records receive FAIL under the study contracts.
Independent human-referenced estimates are 44.2\%--51.6\% on the same sets.
Selected implementation-FAIL evidence is more frequent than semantic-FAIL evidence.
When checked outputs are identical, output comparison cannot distinguish compliant computation from forbidden delegation at any tolerance.
\end{rqanswer}

\section{RQ2: Axis Attribution and Oracle Validity}
\label{sec:rq2}

\subsection{Implementation Axis: Two Complementary Evidence Channels}
\label{subsec:impl-channels}

Of 313 selected implementation FAIL decisions, 299 use only the dispatcher
probe, 6 only launch provenance, and 8 both (Appendix~\ref{s:cases}). On the main corpora,
229/268 FAIL-certificate records carry selected implementation FAIL evidence
(85.4\%).
Eight dispatcher misses involve GEMV work under-counting and unknown-operation
paths (Table~\ref{tab:controls}; Section~\ref{subsec:impl-tier}). The 5\%/30\%
targeting shares do not guarantee detection at those fractions of actual delegated work.

\begin{table}[tb]
\caption{Constructed behavioral controls under fixed audit policies (Appendices~\ref{s:cases}--\ref{s:validity}).
Sealing before result inspection and separation from threshold development are unverified;
capture does not establish population recall or specificity.}
\label{tab:controls}
\footnotesize
\setlength{\tabcolsep}{3pt}
\begin{tabular}{@{}llrp{0.47\linewidth}@{}}
\toprule
Control family & Surface & Result & Misses / notes \\
\midrule
Delegation probes & Impl. & 52/60 & semantic axis adds 4 (56/60 overall);
4 joint misses confirmed by counterfactual replay \\
Launch battery (1 pos., 3 neg.) & Launch & 1/1; 3/3 & the positive
is a below-dispatcher cuBLAS call; no launch flag on negatives;
mediated calls remain subject to Channel 1 \\
Challenge positives & Sem. & 10/12 & misses: stale cache and shape specialization; zero \inconcl{} \\
Challenge negatives & Sem. & 4/4 & no false conviction \\
Timing-tamper probes & Meas. & 2/2 & attributable mutations can convict \\
Hidden-stream probe & Meas. & 0/1 & undetected; guard can only withhold \\
\bottomrule
\end{tabular}
\end{table}

\paragraph{Case 1: Provenance and Contract Interpretation.}
In robust-kbench, the handwritten \texttt{llama\_ffw} passes all 16
semantic policies with an empty dispatcher trace. Vendor launches account for
99.6\% of observed GPU time; all six weak verifiers accept. Our strict
self-implementation policy yields \cthreeb{}. Both annotators label it clean
because ``unchanged GEMMs'' declares the library work their rubric permits. This case is one of
seven reported policy disagreements affecting specificity despite agreement on
execution (Section~\ref{sec:experimental-design}; Cases 2--3 in Appendix~\ref{s:cases}).

\paragraph{Case 4: Withheld Certification and an Independent Semantic Miss.}
\label{subsec:policy-v31}
On \texttt{cuda\_l1\_\_L2\_T80}, the reference is identically zero.
The candidate replaces its class-level \texttt{\_\_call\_\_} with a cached-zero
return; initial outputs match without observed candidate GPU work.
R-B stops the certificate-bound execution before semantic testing, yielding
\inconcl{}. Selected semantic PASS cites a separate record with 16 passing
policies. Independent re-execution finds a new instance returning an
uninitialized \texttt{None} cache instead of the reference tensor.
This witness supplies the human label without entering automatic verdicts:
the audit misses the defect while withholding certification (Appendix~\ref{s:annotation}).
R-B cannot erase an established FAIL.

\subsection{Measurement Evidence and Timing Credibility}
\label{subsec:measurement-axis-results}

The 716-record audit detects no measurement violation, but control results limit this finding.
Timing-primitive checks detect both tampering controls.
The stream guard misses its positive control and can only withhold certification (Table~\ref{tab:controls}).
Selected measurement sources contain guard observations for 672 records (651 PASS, 21 \inconcl{}), leaving 44 without observations.
Each PASS has four required guard PASS results (Appendix~\ref{s:measurement}).
The fixed timing source monitors 577 records with no detected primitive mutation.
The recorded stream ratios reach 1.75 against a 3.0 threshold.
A/A withholding affects speedup reportability, as examined in Section~\ref{sec:rq3}.

\subsection{Validity Against an Independent Blind Reference}
\label{subsec:oracle-validity}

\begin{table}[tb]
\caption{Oracle agreement with the independent blind re-execution
rubric ($n{=}120$): design-weighted estimates and 95\% model-based
posterior intervals (Appendix~\ref{s:annotation}).}
\label{tab:oracle-validity}
\footnotesize
\setlength{\tabcolsep}{3pt}
\begin{tabular}{@{}lr@{\hspace{10pt}}lr@{\hspace{10pt}}lr@{}}
\toprule
Metric & Est.\ [95\% interval] & Metric & Est.\ [95\% interval] & Metric & Est.\
[95\% interval] \\
\midrule
Sensitivity & 0.964 [0.917, 0.967] & Specificity & 0.626 [0.518,
0.673] & PPV & 0.763 [0.638, 0.810] \\
NPV & 0.934 [0.848, 0.943] & Unknown & 3.9\% [3.9, 11.2] &
\inconcl{} defect & 39.5\% [29.3, 57.8] \\
\bottomrule
\end{tabular}
\end{table}

Independent blind review assesses agreement under a separate human rubric.
Two annotators independently inspect artifacts and re-executions without oracle or verifier decisions.
Disagreements on 29/120 units require third-party arbitration.
The rubric permits explicitly declared library computation, so disagreements combine policy interpretation and execution judgments (Case 1).
The confusion matrix cannot isolate detection errors under a shared contract.
Delegation agreement is 80.0\% ($\kappa{=}0.582$), and the reference remains fallible.
Sampling draws 45/45/30 PASS/FAIL/INCONCL records from certificate$\times$corpus strata.
Inclusion probabilities determine the weights.
Evaluated certificates define 47/54/19 metric cells (Appendix~\ref{s:annotation}).

Metric denominators separate decided judgments from unresolved evidence.
The four classification metrics use human-known, oracle-decided records.
Unknown labels and \inconcl{} are reported separately.
The INCONCL-defect share includes the entire cell, including unknowns.
Table~\ref{tab:oracle-validity} describes overall OV120, not main-corpus-only specificity.

Posterior intervals assume shared label probabilities across certificate cells within each design stratum.
The intervals condition on adjudicated labels and exclude annotation and contract-interpretation uncertainty.
Sensitivity to the exchangeability assumption remains untested (Appendix~\ref{s:annotation}).

\subsection{Comparison with Runtime Attribution}
\label{subsec:atrex-comparison}

We compare implementation verdicts with an \emph{Atrex-based attribution screen}
on the same 120 records (110 natural, 10 constructed). Implementation-only
human labels are 34 violations, 80 negatives, and 6 unknowns, distinct from the
full-certificate labels above.

\paragraph{Baseline and Protocol.}
The screen flags suspected delegation when summed matched-kernel device time per forward divided by independently measured median end-to-end latency is $\leq0.5$.
This screening rule is not an Atrex verdict and adds no causal or contract test.
We reuse pinned Atrex~\cite{yang2026llm} timing, ratio, and internal-error modules, adapting task/state initialization, exact CUDA/Triton symbol registration, and outer exceptions (Appendix~\ref{s:annotation}).
Attribution is unavailable for 9 records, 6 with known human labels.

\begin{table}[tb]
\caption{Implementation comparison on the 114 records with known human
delegation labels; 6 human-unknown records are excluded. Positives denote undeclared delegation, predicted from low attribution or
implementation FAIL. Coverage counts decisions; abstentions are neither clean
nor detected violations.}
\label{tab:atrex-comparison}
\footnotesize
\setlength{\tabcolsep}{5pt}
\begin{tabular}{@{}lrrrrrr@{}}
\toprule
Method & TP & FP & FN & TN & Abstain & Coverage \\
\midrule
Our implementation axis & 29 & 14 & 2 & 55 & 14 & 87.7\% \\
Atrex-based screen & 33 & 48 & 0 & 27 & 6 & 94.7\% \\
\bottomrule
\end{tabular}
\end{table}

\paragraph{Errors and Coverage.}
The oracle reduces false positives at the cost of detection and coverage (Table~\ref{tab:atrex-comparison}).
The implementation axis detects 29/34 human-positive records versus 33/34 for the screen, and decides 100/114 known-label records versus 108/114.
Among 98 jointly decided records, only the oracle agrees with the rubric on 29 and only the screen on 2 (controls missed by targeting).
These comparisons remain sample-conditional.
Low attribution alone cannot establish forbidden work because end-to-end latency includes time outside observed target kernels (Appendix~\ref{s:annotation}).

\subsection{\inconcl{}: Unresolved Evidence and Decidability}
\label{subsec:inconclusive}

The ledger contains 147 \inconcl{} certificates (20.5\%), while 286 records
have 445 unresolved selected axes (diagnostic categories in Appendix~\ref{s:inconcl}).
These axis labels neither partition certificate causes nor assign responsibility.
An unresolved channel does not erase another channel's independent FAIL evidence.
Eight of the 19 sampled \inconcl{} records have human defect labels.
Dispatcher probes likewise yield implementation FAIL on 17 executed PyTorch
fallback paths among 46 CUDA-L1 build-failure candidates (Appendix~\ref{s:cases}).

\begin{rqanswer}\textbf{Answer to RQ2.}
Against the blind reference, design-weighted sensitivity is 0.964, NPV 0.934, and specificity 0.626 on decided, human-known records.
The implementation axis has fewer false positives than the attribution screen, with lower detection and coverage.
Labels remain policy-relative, and Case 4 demonstrates a missed semantic defect despite withheld certification.
PASS denotes no detected violation on audited surfaces.
\end{rqanswer}

\section{RQ3: Credibility and Certification of Speedup Claims}
\label{sec:rq3}

RQ3 reports two descriptive quantities: the \emph{claim-to-local contrast},
the median record-wise ratio $m_i/c_i$ of reportable local speedup to upstream
claim; and \emph{certificate survival}, the paired cohort's PASS fraction.
Pairs link corresponding candidates across settings, with reference, input,
and state conditions not uniformly matched. The contrast neither isolates
hardware or toolchain effects nor establishes upstream claim overstatement.

Of 577 records with A/A results, 394 speedups are withheld and 183 records pass.
Three passes lack positive recorded speedups, leaving 180 \emph{measurable} records (Section~\ref{subsec:meas-tier}; details in Appendix~\ref{s:measurement}).
Across all seven corpora, these 180 records receive 32 PASS, 121 FAIL, and 27 \inconcl{} certificates.
Among the measurable records, 64 CUDA-L1 and 68 AI-CUDA-Engineer records with upstream claims form the 132-record main paired cohort.
A/A withholding affects speedup reportability, not certificate status.

\begin{table}[tb]
\caption{Paired speedups across upstream/local evaluation settings and certificate
survival; med.($m/c$) is the median of within-record ratios.
The last column gives PASS$/n$ and the PASS subset's claimed and measured
medians. The TritonBench-G row has no upstream claims.
PASS denotes no detected violation on the audited surfaces.}
\label{tab:speedup-evaporation}
\footnotesize
\setlength{\tabcolsep}{3.0pt}
\begin{tabular}{@{}lrrrrr@{}}
\toprule
Corpus and timing cohort & $n$ & Claimed med.\ & Measured med.\ &
med.($m/c$) & PASS (med.\ $c$, med.\ $m$) \\
\midrule
CUDA-L1 & 64 & 1.413$\times$ & 1.130$\times$ & 0.920 &
0 / 64 \\
AI-CUDA-Engineer & 68 & 1.143$\times$ & 1.051$\times$ & 0.958 &
22 / 68\ \ (1.146$\times$, 1.505$\times$) \\
Main pooled & 132 & 1.283$\times$ & 1.094$\times$ & 0.935 &
22 / 132 (1.146$\times$, 1.505$\times$) \\
\midrule
TritonBench-G self-built (no claims) & 22 & --- &
1.793$\times$ & --- & 8 / 22\ \ (---, 1.647$\times$) \\
\bottomrule
\end{tabular}
\end{table}

Most records in the paired main cohort remain uncertified (Table~\ref{tab:speedup-evaporation}).
Of 132 records, 22 receive PASS, 93 receive FAIL, and 17 remain \inconcl{}.
CUDA-L1's two PASS certificates lie outside its 64-record paired cohort.

Implementation violations account for most uncertified records under the semantic $>$ implementation $>$ measurement priority (Section~\ref{sec:experimental-design}).
Using certificate-bound complete records, the 110 uncertified records comprise 24 (21.8\%) semantic violations, 69 (62.7\%) implementation violations, 0 measurement violations, and 17 (15.5\%) unresolved cases.
Implementation assignments exclude semantic overlaps, and unresolved cases are not counted as violations.

Certificate evidence does not establish that violations caused recorded speedups.
All 22 PASS certificates share complete records with their timings, but 54 uncertified records (37 FAIL, 17 \inconcl{}) use different sources.
Candidate code is linked across these sources, but full input/state equality is not established for every pair (Appendix~\ref{s:measurement}).

Certification does not require a speed advantage.
Li et al.~\cite{li2026correctslow} independently examine the gap between
numerical correctness and performance as a replacement for optimized library operators.
The PASS-subset medians in Table~\ref{tab:speedup-evaporation} describe selected records, not an audit-induced improvement.
In TritonBench-G's self-built layer, 8/22 measurable records receive PASS, including records at $0.19\times$ and $0.87\times$.

Timing eligibility limits the scope of the paired summaries.
A/A passage is associated with reference duration on the shared, unlocked-clock host (Appendix~\ref{s:measurement}).
The association does not isolate workload, measurement overhead, or clock effects.
Main paired summaries require reportable timings and upstream claims, while PASS-subset medians additionally condition on certification (Section~\ref{sec:threats}).

\begin{rqanswer}\textbf{Answer to RQ3.}
On the paired main-corpus cohort, the median measured/claimed ratio is 0.935 and certificate survival is 22/132 (16.7\%).
Implementation violations account for 69 of 110 uncertified records under the stated priority, while 17 remain inconclusive.
These are conditional record-level results, and certification does not guarantee a speed advantage.
\end{rqanswer}

\section{RQ4: A Training-Free Gate Swap}
\label{sec:selector-intervention}

RQ4 compares shipping rules over fixed candidates and saved evidence,
without retraining, resampling, or candidate repair.

\paragraph{Design.}
\looseness=-1
The pool contains 232 tasks $\times$ 5 slots $=$ 1{,}160 candidates:
206 CUDA tasks selected by the resource plan from 220 joined tasks,
and all 26 joined TritonBench tasks (Appendix~\ref{s:rq4}).
Slots follow CUDA-L1, AI-CUDA-Engineer, then Opus c0--c2 for CUDA;
TritonBench uses G, pref, then Opus c0--c2.
All five share the first registered source's reference.
Unavailable evidence remains \inconcl{}; all slots and tasks stay in the denominators.

We reconcile 204 slots with complete primary-audit records after matching
candidate source, task, reference, and policy, including verified state adapters.
The other 956 retain historical RQ4 evidence; this reconciliation reuses evidence
without a new GPU run of the pool. Certificates never combine axes from different records.
The oracle-visible $2{\times}2$ comparison crosses eligibility with first/fastest ordering.
$S_0$ and $S_0$-fast require the corresponding frozen weak verifier's PASS.
$S_1$-first and $S_1$ require PASS in the reconciled eligibility ledger.
For reused records, eligibility requires the complete-record PASS certificate; historical slots
retain saved three-axis PASS plus R-B. Axis-only $S_{\mathrm{axis}}$-first
and $S_{\mathrm{axis}}$ instead require three-axis PASS.
The historical evidence lacks some current certification checks, so this
mixed-evidence study does not evaluate full certification on every slot.

Fastest selects the largest usable cached local speedup, breaking ties by fixed order.
Speedups require measurement PASS, inactive guards, and A/A and paired A/B checks;
changed adapters cannot inherit timing without revalidation (Appendix~\ref{s:rq4}).
No new timing is measured. If no eligible slot has timing, the first is selected;
if none is eligible, the selector abstains.
Of $S_1$'s 62 deliveries, 15 use timing and 47 use fallback.

Human labels do not enter selection. The historical 330-unit blind census
covers all but one candidate in the refreshed 331-unit delivery union (Appendix~\ref{s:annotation}).
The two annotators used Section~\ref{subsec:oracle-validity}'s rubric;
packets hid selector membership. There were 73 arbitrated units;
delegation agreement was 81.2\% ($\kappa{=}0.549$).
The refresh follows that review and does not claim a new prospective blind study.
We retain the missing label as unknown and report unsafe/232,
unsafe/shipments, and coverage; abstention is not successful delivery.

\paragraph{Batch Timing Attribution.}
\label{subsec:rq4-results}
\looseness=-1
Isolated execution does not substantiate candidate attribution for the 114 historical batch timing signals under R-C: ratios have median 1.000 and maximum 1.046 (Appendix~\ref{s:rq4}).
Their speedups remain unavailable for ranking.

\paragraph{Evidence Availability Limits Coverage.}
The reconciled pool has 184 eligible, 789 inconclusive, and 187 FAIL slots.
Eligible slots occur on 62 tasks, bounding $S_1$ coverage at
62/232 (26.7\%) regardless of ordering, versus $S_0$'s 40.1\%--90.5\%.

\begin{table}[tb]
\caption{Gate-swap results on 232 tasks. Oracle-visible unsafe is
\ctwo{}$\cup$C3$*$; human unsafe uses independent labels.
Rates divide by 232 except parenthesized human unsafe/shipped.
Lower-panel bounds assign the one unknown shipment clean or unsafe;
they are identification bounds, not confidence intervals.}
\label{tab:selector-results}
\footnotesize
\setlength{\tabcolsep}{3.4pt}
\begin{tabular}{@{}lrrrrr@{}}
\toprule
Weak verifier & $S_0$ unsafe & $S_0$ C3$*$ & $S_0$/$S_0$-fast cov. &
$S_0$-fast unsafe & Human $S_0$ unsafe \\
\midrule
CUDA-Agent & 11.6\% (27) & 9.5\% (22) & 93/232 & 9.1\% (21) & 6.0\% (15.1\%) \\
KernelBench & 39.7\% (92) & 37.1\% (86) & 183/232 & 38.8\% (90) & 27.6\% (35.0\%) \\
KernelGYM & 48.7\% (113) & 45.3\% (105) & 210/232 & 47.0\% (109) & 31.5\% (34.8\%) \\
robust-kbench & 40.9\% (95) & 39.7\% (92) & 173/232 & 39.7\% (92) & 30.2\% (40.5\%) \\
DeepReinforce & 28.4\% (66) & 24.1\% (56) & 171/232 & 25.9\% (60) & 22.4\% (30.4\%) \\
MusaCoder & 21.1\% (49) & 17.7\% (41) & 167/232 & 20.3\% (47) & 21.6\% (29.9\%) \\
\bottomrule
\end{tabular}
\par\smallskip
\begin{tabular}{@{}lrrr@{}}
\toprule
Audit selector & Coverage & Human unsafe/232 & Human unsafe/shipped \\
\midrule
Axis-only $S_{\mathrm{axis}}$ & 63/232 (27.2\%) & [1,2]/232 & [1,2]/63 (1.6--3.2\%) \\
RQ4 gate $S_1$ & 62/232 (26.7\%) & [1,2]/232 & [1,2]/62 (1.6--3.2\%) \\
\bottomrule
\end{tabular}
\end{table}

\paragraph{Oracle-Visible Gate and Ranking Effects.}
\looseness=-1
Eligibility has a larger effect than ordering on oracle-visible unsafe shipments.
Contrasts average over the other factor and equally over six verifiers:
\textbf{30.9 pp} for eligibility (range 10.3--47.8) and 0.8 pp for ordering (Appendix~\ref{s:rq4}).
Both audit-gated cells have zero unsafe shipments by definition;
these contrasts do not independently validate the audit labels.

\paragraph{Human-Labeled Shipment Outcomes.}
\looseness=-1
$S_1$ ships 60 candidates labeled clean, one labeled unsafe, and one without a label.
The known unsafe shipment is CUDA-L1 L2 T14 (BR-203), labeled undeclared
delegation despite its reused PASS certificate. The unreviewed shipment is
TritonBench \texttt{109\_softmax\_optimize}, slot 1.
Thus human-unsafe deliveries lie between 1 and 2, or 1.6\%--3.2\% of shipments
and 0.4\%--0.9\% of tasks. These bounds remain below the six $S_0$ rates,
while eligibility, ordering, and abstention change jointly.
We withhold paired human $p$-values until the missing label is resolved;
the previous census's tests do not describe the refreshed selections.
The mixed-evidence gate does not establish zero unsafe delivery.

\paragraph{Delivery with Speedup.}
SafeFast@$k$ is the share of all 232 tasks shipping an $S_1$-eligible
candidate with credible speedup $>k\times$.
Changing first to fastest raises SafeFast@$1$ from 4/232 to 6/232 (2.6\%),
while SafeFast@$2$ stays at 2/232 (0.9\%).
$S_0$-fast achieves SafeFast@$1$ on 3--5 tasks and SafeFast@$2$ on two (Appendix~\ref{s:rq4}).

\paragraph{Cost Diagnostics.}
Historical cache-write intervals provide indirect diagnostics for one 46-task segment.
Strong-/weak-cache p50s are 7.460 min and 77.142\,s ($\approx5.8\times$; Appendix~\ref{s:rq4}).
These unpaired, differently filtered intervals do not measure the refresh,
chosen-candidate post-filtering, net additional latency, or full-pool screening cost.

\begin{rqanswer}\textbf{Answer to RQ4.}
Audit-based selection lowers human-unsafe delivery in this fixed pool alongside lower coverage.
$S_1$ covers 26.7\% of tasks, with one known unsafe and one unknown among
62 shipments: the unsafe fraction is bounded by 1.6\%--3.2\%.
\end{rqanswer}

\section{Threats to Validity}
\label{sec:threats}

\paragraph{Internal.}
\looseness=-1
RQ4's shared-process timing signals require isolated attribution checks;
unresolved channels cannot support conviction
(Section~\ref{subsec:rq4-results}). With unlocked clocks, ABBA and A/A
stability do not establish unbiased timing or drift bounds.
RQ3 conditions on reportable timings and upstream claims, plus certification
for PASS-subset summaries. Unmatched reference/input/state conditions preclude
isolating environmental effects or attributing speedup to violations from these
contrasts alone (Section~\ref{sec:rq3}).

\paragraph{Construct.}
\looseness=-1
Labels concern policy-relative decisions, not intent; permitted fixed-shape
optimization alone cannot convict. PASS means no detected violation on
audited surfaces. Measurement conviction uses only positively controlled
timing-primitive checks; runtime precision-mode evidence is unimplemented.
Dispatcher checks do not establish control-flow equivalence.
The sealed launch inventory covers cuBLAS/cuDNN only. Tests cover finite
inputs and call sequences; adaptive evasion remains unevaluated.
R-A requires observed candidate work; R-B withholds certification for
listed degenerate-reference tasks. Neither prerequisite establishes a violation.
Detection at the fixed 5\%/30\% targeting shares depends on work
proxies and supported operations (Section~\ref{subsec:impl-channels}).
The audit and human rubrics disagree on declared delegation in Case 1;
agreement metrics do not isolate detection errors (Appendix~\ref{s:annotation}).
OV120 validity intervals share label probabilities across certificate cells
within each original sampling stratum. Exchangeability sensitivity remains
untested; intervals fix adjudicated labels and do not quantify
annotation or contract-interpretation uncertainty (Section~\ref{subsec:oracle-validity}).

\paragraph{External and Statistical Conclusion.}
The audit of 716 records characterizes the evaluated adapters. Its results do not
generalize to generator populations or other GPU architectures.
DeepReinforce and MusaCoder use publication-derived
reconstructions; the Atrex-based comparison covers implementation decisions on
120 records, with attribution availability dependent on the environment.
Core self-implementation is our study policy, not a uniform upstream requirement.
TritonBench source reuse
can violate McNemar's independent-pair assumption
(Section~\ref{sec:experimental-design}). Protocols are not externally preregistered.

RQ4's fixed, resource-planned pool contains 1{,}160 slots across 232 tasks.
Its reconciled gate covers 62 tasks, with one known unsafe and one unknown
shipment; unsafe/shipments is bounded by 1.6\%--3.2\%.
The 204 slots with reused complete evidence and 956 historical slots retain different
evidence scopes; this refresh is not a uniform full-certification rerun.
The refresh follows the original blind census, whose 330 labels do not cover
one new delivery. Paired human tests are withheld, and lower unsafe delivery
does not hold coverage fixed. Prospective risk and deployment overhead
remain unevaluated (Section~\ref{sec:selector-intervention}).

\section{Related Work}
\label{sec:related-work}

\paragraph{LLM-Generated GPU Kernels and Validators.}
\looseness=-1
KernelBench~\cite{ouyang2025kernelbench} and TritonBench~\cite{li2025tritonbench} established execution-based evaluation for LLM-written CUDA/Triton kernels; robust-kbench~\cite{lange2025towards}, KernelGYM~\cite{liu2026dr}, CUDA-Agent~\cite{dai2026cuda}, MusaCoder~\cite{cheng2026musacoder}, and DeepReinforce~\cite{deepreinforce} add stronger evaluation, RL environments, fallback checks, and timing-attack defenses.
GEAK~\cite{wang2025geak}, the AI CUDA Engineer~\cite{lange2025ai}, CUDA-L1~\cite{sun2026cuda}, and KernelBenchX~\cite{wang2026kernelbenchx} study agentic and RL optimization.
Recent work~\cite{sarkar2026correctness,zhang2026kernelbench,martinez2026next,shui2026harness,veit2026kernel,shah2026contract} adds fp64-backed fuzzing, hidden tests, degenerate-task exclusion, verified solutions, harness engineering, contract languages, and a twelve-gate audit of 2{,}638 accepted kernels; our results show why output checks alone leave implementation violations unresolved.

Atrex's production-derived benchmark~\cite{yang2026llm} is the closest runtime-attribution comparison: its target-DSL time ratio exposes fallback use beyond correctness.
Section~\ref{subsec:atrex-comparison} tests that signal against independent delegation labels; our combined implementation axis has fewer false positives under that rubric, with lower detection and coverage.
Formal verification targets race/barrier safety~\cite{betts2012gpuverify} or semantic equivalence~\cite{driscoll2025equivalence} rather than reward validators.
Our intervention builds on contract vocabulary, semantic re-auditing, runtime attribution, causal testing, fault injection, and dynamic slicing~\cite{johnson2020causal,jia2010analysis,agrawal1990dynamic} to audit six verifiers, account for abstention, and evaluate a shipping gate.

\paragraph{GPU Defects and Performance Evaluation.}
Studies of tile-program code-generation bugs~\cite{rathnasuriya2026tilebugs}
and GPU numerical bugs~\cite{rathnasuriya2025gpunumeric} characterize
input-sensitive failures, their causes, and detection strategies; the latter
also presents GPU-NBDetect.
Li et al.~\cite{li2026correctslow} find that numerically valid Triton and
TileLang kernels can remain much slower than optimized library operators,
and examine library-relative efficiency and roofline-based checks.
These studies complement our focus on validators' contract-dependent
acceptance decisions and the credibility of their timing evidence.

\paragraph{Executable Code Evaluation and Oracle Design.}
\looseness=-1
CodeRL~\cite{le2022coderl} uses execution feedback.
EvalPlus~\cite{liu2023your} exposes inadequate tests.
Oracle design~\cite{barr2014oracle} and metamorphic testing~\cite{chen2018metamorphic} address the reliability of such judgments.
GPU rewards additionally require evidence about implementation and measurement.
Reward hacking~\cite{amodei2016concrete,skalse2022defining} and learned verifier exploitation~\cite{ray2026before} motivate auditing these non-output obligations before deploying a validator as a gate.

\paragraph{Model Uncertainty and Selective Prediction.}
CodeImprove~\cite{rathnasuriya2025codeimprove} uses a validity score to identify
inputs beyond a code model's effective scope and adapts them through
semantics-preserving program transformations.
For LLM-based code generation, Rathnasuriya and Yang~\cite{rathnasuriya2026overconfident}
report that confidence and entropy provide model-dependent evidence of
execution failure, and uncertainty-based selection does not consistently
improve accepted-set accuracy.
Their study of defect prediction and vulnerability detection~\cite{rathnasuriya2026calibration}
also finds that improved post-hoc calibration scores need not improve
selective prediction.
These approaches assess model-derived reliability signals.
In \systemname{}, \inconcl{} instead records unmet auditability or certification
conditions; RQ4 evaluates unsafe delivery together with coverage under those
evidence constraints.

\section{Conclusion}
\label{sec:conclusion}

In this paper, we have contributed a contract model and an audit oracle for execution-based validators
in LLM-driven GPU-kernel generation. The model separates semantic, implementation,
and measurement obligations. The oracle combines reference-calibrated semantic
tests, complementary dispatcher interventions and launch provenance, and timing
guards, with explicit abstention for unresolved checks. A cross-verifier audit
and independent human evaluation assess validator blind spots and audit validity
under the study policies. A separate fixed-pool study examines how eligibility
and ordering affect unsafe delivery and coverage. Together, these contributions
support testing reward and shipping decisions against task obligations
and execution evidence on audited surfaces.
\raggedbottom
\par\medskip\noindent
\begin{minipage}{\linewidth}
\begin{acks}
This work was partially supported by the Fundamental and Interdisciplinary Disciplines Breakthrough Plan of the Ministry of Education of China (No.~JYB2025XDXM118) and the National Natural Science Foundation of China under Grant Nos.~U25A6023 and 92464301.
Tao Xie is also affiliated with Key Laboratory of High Confidence Software Technologies (Peking University), Ministry of Education, China; School of Computer Science, Peking University, China; Institute of Systems for Advanced Computing, Fudan University, China; Shanghai Institute of Systems for Open Computing, China; School of Computer Science and Engineering, Southeast University, China.
\end{acks}

\end{minipage}
\par

\clearpage
\bibliographystyle{ACM-Reference-Format}
\bibliography{references}

\clearpage
\appendix
\raggedbottom
\counterwithin{table}{section}
\counterwithin{figure}{section}
\counterwithin{equation}{section}
\counterwithin{algorithm}{section}
\setcounter{table}{0}
\setcounter{figure}{0}
\setcounter{equation}{0}
\setcounter{algorithm}{0}

\renewcommand{\theHtable}{appendix.\Alph{section}.\arabic{table}}
\renewcommand{\theHfigure}{appendix.\Alph{section}.\arabic{figure}}
\renewcommand{\theHequation}{appendix.\Alph{section}.\arabic{equation}}
\providecommand{\theHalgorithm}{}
\renewcommand{\theHalgorithm}{appendix.\Alph{section}.\arabic{algorithm}}

\providecommand{\mainpapertableref}[2]{Table~\ref{#1}}
\providecommand{\mainpapersectionref}[2]{Section~\ref{#1}}

\section*{Appendices}
Appendices~\ref{s:params}--\ref{s:annotation} provide audit rules, supporting results, and evidence
boundaries for the main paper. Each analysis states its estimand and uses
the corresponding sealed records or completed annotation sheets.
Summaries and RQ3 claimed/measured speedup pairs are in
\texttt{experimental\_tables.json}; \texttt{final\_decisions.json} lists
certificate and selected-axis sources, and \texttt{cause\_inventory.json}
records unresolved-axis evidence labels. \texttt{comparison\_verified.json}
retains OV120 sampling fields and the implementation-comparison results.

\section{Audit Parameters and Record Provenance}
\label{s:params}

The verdict semantics, certificate aggregation, and outcome-code priority
are defined in \mainpapersectionref{subsec:verdict-semantics}{3.1},
\mainpapersectionref{subsec:conservative-aggregation}{3.5}, and
\mainpapertableref{tab:outcome-taxonomy}{1}.
This appendix gives the source-handling and replay details behind those rules.

\paragraph{Task-Policy Resolution.}
The registry-to-task loader prioritizes explicit
boolean implementation policies, uses the manifest for missing/null policies,
and rejects invalid values or uncovered entries.

\paragraph{Source Qualification and Selection.}
Qualification checks file and progress hashes, record schema
and digests, record membership, and manifest-bound run or experiment identifiers.
Ordinary replays preserve record ID and task, candidate, contract, and policy
fingerprints. Declared state adaptations require separate verification of
candidate/reference sources, registration, adaptation fields, and adapted
candidate-fingerprint correspondence.
Worker placeholders require a registry-entry digest, the same record ID, and
\inconcl{} on all axes and the certificate. They record unavailable evidence,
without establishing complete state correspondence or supplying PASS/FAIL evidence.
Admission across records does not require identical inputs, environments, or
state hashes; state correspondence remains necessary within semantic comparisons.

Source export traverses records in specified input order, applying certificate
eligibility before selecting each axis and the effective certificate separately.
Restricted merging admits new decisions only for unresolved evidence; axis
updates additionally require a pre-specified eligibility set. Let $D$ denote
PASS or FAIL and $U$ denote \inconcl{}. The operations differ as follows:

\begin{center}
\footnotesize
\setlength{\tabcolsep}{3pt}
\begin{tabular}{@{}p{0.24\linewidth}p{0.32\linewidth}p{0.37\linewidth}@{}}
\toprule
Existing / incoming & Source export & Restricted merge \\
\midrule
$D/U$ & Keep existing source & Keep existing source \\
$U/D$ & Use incoming source & Update an eligible axis; a certificate requires an incoming record's decided certificate \\
Same status & Use incoming source & Keep existing source \\
Opposite decided statuses & Use incoming source; no conflict guard in this operation & Stop import for conflict investigation \\
\bottomrule
\end{tabular}
\end{center}

Restricted imports require empty weak results and preserve existing
weak-evidence sources. Input order, rather than a timestamp ranking,
determines replacement during export. The example in Appendix~\ref{s:inconcl}
distinguishes certificate eligibility from source replacement.

Source digests identify the base policy and vendor inventory. Executed
settings additionally depend on each record's \texttt{effective\_policy},
tolerance configuration, and timing fields. The inventory's internal
fingerprint agrees with the bound execution records.

\paragraph{Dispatcher Replay Checks.}
Beyond the shape, dtype, and structure checks in
\mainpapersectionref{subsec:impl-tier}{3.3}, recursive comparisons check
tensor/non-tensor types, mapping keys, and sequence types/lengths.
Matched call counts are not required.

\paragraph{Stress-Schedule Configuration.}
The base schedule combines 12 pre-specified slots with four native-seed
slots and uses contract-adapted values spanning $10^{-4}$--$10^{5}$.
Appendix~\ref{s:miss-matrix} gives the instantiated provider-specific schedule.

\paragraph{A/A Resampling Details.}
For the timing gate in \mainpapersectionref{subsec:meas-tier}{3.4},
let $t_{abi}$ be the $i$th latency of arm $a$ in batch $b$ and
$m_{ab}=\operatorname{median}_{1\leq i\leq20}t_{abi}$.
The sample standard deviation uses denominator $60-1$.
Let $\mathcal P$ contain the odd/even and first-ten/last-ten position pairs
within each arm and batch. The order-effect requirement is
\[
 O=\max_{a,b}\max_{(I,J)\in\mathcal P}
 \frac{\left|\operatorname{median}_{i\in I}t_{abi}
             -\operatorname{median}_{i\in J}t_{abi}\right|}{m_{ab}}
 \leq0.05.
\]
The three paired batch medians are resampled with replacement 2{,}000 times
(seed 0), using identical batch indices for both arms. Each replicate takes
the median of the resampled A-arm medians divided by that of the B arm.
The 2.5th and 97.5th percentiles provide the interval $[L,U]$ used by
the main-text acceptance rule. The three-batch construction does not
guarantee nominal coverage.

\section{Per-Corpus Misses and Aggregate False Rejects}
\label{s:miss-matrix}

The rate definitions and weak-REJECT convention follow
\mainpapersectionref{sec:experimental-design}{4}.
The following details specify corpus pooling, unresolved-record bounds,
and paired verifier contrasts.

\paragraph{Pooling and Abstention Sensitivity.}
For a fixed verifier, let $a_c$ count accepted FAIL certificates and $b_c$
count decided accepts in corpus $c$. For $\sum_c b_c>0$, the pooled rate is
\[
\mathrm{FA}_{\mathrm{pool}}
=\frac{\sum_c a_c}{\sum_c b_c}
=\sum_{c:b_c>0}\frac{b_c}{\sum_j b_j}\frac{a_c}{b_c}.
\]
Weights depend on each verifier's decided acceptance counts.
With $a=\sum_c a_c$, $b=\sum_c b_c$, and $u$ accepted \inconcl{}
records, the sensitivity endpoints are $a/(b+u)$ and $(a+u)/(b+u)$
for $b+u>0$. They assign all $u$ records to PASS or FAIL with weak decisions
and resolved certificates fixed; they quantify assignment sensitivity,
not sampling uncertainty or errors in resolved certificates.

\begin{table}[htbp]
\caption{Misses (accepted oracle-FAILs / corpus FAIL total) and false
rejects (rejected oracle-PASSes / corpus PASS total).}
\label{tab:appendix-02}
\scriptsize
\setlength{\tabcolsep}{3.2pt}
\begin{tabular}{@{}lrrrrrr@{}}
\toprule
Corpus & CUDA-Agent & KernelBench & KernelGYM & robust-kb. & DeepReinf. &
MusaCoder \\
\midrule
CUDA-L1 miss & 23/148 & 117/148 & 136/148 & 110/148 & 75/148 & 39/148 \\
AI-CUDA-Eng.\ miss & 97/120 & 86/120 & 97/120 & 83/120 & 98/120 & 98/120 \\
KernelBenchX miss & 17/32 & 21/32 & 23/32 & 20/32 & 23/32 & 16/32 \\
TB-G self-built miss & 8/21 & 15/21 & 15/21 & 10/21 & 14/21 & 12/21 \\
TB-G upstream miss & 1/17 & 1/17 & 1/17 & 1/17 & 1/17 & 1/17 \\
Delegation controls miss & 55/56 & 56/56 & 56/56 & 39/56 & 56/56 &
\textbf{1/56} \\
robust-kbench miss & 2/3 & 2/3 & 2/3 & 2/3 & 2/3 & 2/3 \\
\midrule
All-corpus false rejects & 15/172 & 17/172 & 9/172 & 21/172 & 11/172 &
14/172 \\
\bottomrule
\end{tabular}
\end{table}

\paragraph{Paired Verifier Contrasts.}
McNemar's exact null model assigns probability $1/2$ to either discordance
direction among the $b+c$ discordant FAIL records. With independent records,
$b\sim\mathrm{Binomial}(b+c,1/2)$. Holm correction covers the 15 pairs within
each pool; it does not remove source dependence.

\begin{table}[htbp]
\caption{Paired misses on 397 FAIL-certificate records. $b$/$c$ count misses
unique to the second/first verifier. Exact two-sided $p$ and Holm adjustments
follow the model above; 14 of 15 pairs have adjusted $p<0.05$, with
robust-kbench vs.\ DeepReinforce the exception. Shared source kernels limit independent-record inference.}
\label{tab:appendix-03}
\scriptsize
\setlength{\tabcolsep}{4pt}
\begin{tabular}{@{}lrrrr@{}}
\toprule
Pair & $b$ & $c$ & $p$ (exact) & $p$ (Holm) \\
\midrule
CUDA-Agent vs.\ KernelBench & 112 & 17 & 2.45e-18 & 2.45e-17 \\
CUDA-Agent vs.\ KernelGYM & 127 & 0 & 1.18e-38 & 1.65e-37 \\
CUDA-Agent vs.\ robust-kbench & 106 & 44 & 4.40e-07 & 1.76e-06 \\
CUDA-Agent vs.\ DeepReinforce & 66 & 0 & 2.71e-20 & 2.98e-19 \\
CUDA-Agent vs.\ MusaCoder & 37 & 71 & 1.38e-03 & 4.14e-03 \\
KernelBench vs.\ KernelGYM & 32 & 0 & 4.66e-10 & 2.79e-09 \\
KernelBench vs.\ robust-kbench & 2 & 35 & 1.02e-08 & 5.12e-08 \\
KernelBench vs.\ DeepReinforce & 29 & 58 & 2.48e-03 & 4.96e-03 \\
KernelBench vs.\ MusaCoder & 22 & 151 & 7.43e-25 & 8.91e-24 \\
KernelGYM vs.\ robust-kbench & 2 & 67 & 8.19e-18 & 7.37e-17 \\
KernelGYM vs.\ DeepReinforce & 2 & 63 & 1.16e-16 & 9.31e-16 \\
KernelGYM vs.\ MusaCoder & 2 & 163 & 5.86e-46 & 8.79e-45 \\
robust-kbench vs.\ DeepReinforce & 57 & 53 & 7.75e-01 & 7.75e-01 \\
robust-kbench vs.\ MusaCoder & 31 & 127 & 5.38e-15 & 3.77e-14 \\
DeepReinforce vs.\ MusaCoder & 0 & 100 & 1.58e-30 & 2.05e-29 \\
\bottomrule
\end{tabular}
\end{table}

\begin{table}[htbp]
\caption{Paired misses on the five natural corpora (CUDA-L1,
AI-CUDA-Engineer, KernelBenchX, both TritonBench-G layers; 338 FAIL records).
Delegation controls and the handwritten set are excluded. Under the same
model, correction, and source-dependence limitation, 14 of 15 pairs have
adjusted $p<0.05$. Natural-corpus miss rates: CUDA-Agent
43.2\%, KernelBench 71.0\%, KernelGYM 80.5\%, robust-kbench 66.3\%,
DeepReinforce 62.4\%, MusaCoder 49.1\%.}
\label{tab:appendix-04}
\scriptsize
\setlength{\tabcolsep}{4pt}
\begin{tabular}{@{}lrrrr@{}}
\toprule
Pair & $b$ & $c$ & $p$ (exact) & $p$ (Holm) \\
\midrule
CUDA-Agent vs.\ KernelBench & 111 & 17 & 4.25e-18 & 5.10e-17 \\
CUDA-Agent vs.\ KernelGYM & 126 & 0 & 2.35e-38 & 3.53e-37 \\
CUDA-Agent vs.\ robust-kbench & 105 & 27 & 4.66e-12 & 3.26e-11 \\
CUDA-Agent vs.\ DeepReinforce & 65 & 0 & 5.42e-20 & 7.05e-19 \\
CUDA-Agent vs.\ MusaCoder & 37 & 17 & 9.07e-03 & 1.81e-02 \\
KernelBench vs.\ KernelGYM & 32 & 0 & 4.66e-10 & 2.79e-09 \\
KernelBench vs.\ robust-kbench & 2 & 18 & 4.02e-04 & 1.61e-03 \\
KernelBench vs.\ DeepReinforce & 29 & 58 & 2.48e-03 & 7.44e-03 \\
KernelBench vs.\ MusaCoder & 22 & 96 & 3.26e-12 & 2.61e-11 \\
KernelGYM vs.\ robust-kbench & 2 & 50 & 6.12e-13 & 5.51e-12 \\
KernelGYM vs.\ DeepReinforce & 2 & 63 & 1.16e-16 & 1.28e-15 \\
KernelGYM vs.\ MusaCoder & 2 & 108 & 9.41e-30 & 1.32e-28 \\
robust-kbench vs.\ DeepReinforce & 40 & 53 & 2.13e-01 & 2.13e-01 \\
robust-kbench vs.\ MusaCoder & 31 & 89 & 1.12e-07 & 5.59e-07 \\
DeepReinforce vs.\ MusaCoder & 0 & 45 & 5.68e-14 & 5.68e-13 \\
\bottomrule
\end{tabular}
\end{table}

\begin{table}[htbp]
\caption{Abstention sensitivity by natural corpus. Counts are weak-accepted
\inconcl{} records; bounds assign those records to PASS or FAIL with resolved
certificates fixed. Decided acceptance counts and false accepts are in
\mainpapertableref{tab:main-fa}{6}.}
\label{tab:appendix-05}
\footnotesize
\setlength{\tabcolsep}{5pt}
\begin{tabular}{@{}llrr@{}}
\toprule
Corpus & Verifier & Accepted \inconcl{} & FA bounds\,(\%) \\
\midrule
CUDA-L1 & CUDA-Agent & 17 & [57.5,100.0] \\
 & KernelBench & 41 & [74.1,100.0] \\
 & KernelGYM & 46 & [73.9,98.9] \\
 & robust-kbench & 38 & [74.3,100.0] \\
 & DeepReinforce & 25 & [75.0,100.0] \\
 & MusaCoder & 10 & [79.6,100.0] \\
AI-CUDA-Engineer & CUDA-Agent & 13 & [47.5,53.9] \\
 & KernelBench & 6 & [47.8,51.1] \\
 & KernelGYM & 13 & [47.5,53.9] \\
 & robust-kbench & 5 & [48.3,51.2] \\
 & DeepReinforce & 13 & [47.8,54.1] \\
 & MusaCoder & 13 & [47.8,54.1] \\
KernelBenchX & CUDA-Agent & 2 & [50.0,55.9] \\
 & KernelBench & 2 & [52.5,57.5] \\
 & KernelGYM & 2 & [54.8,59.5] \\
 & robust-kbench & 1 & [52.6,55.3] \\
 & DeepReinforce & 2 & [54.8,59.5] \\
 & MusaCoder & 1 & [48.5,51.5] \\
TritonBench-G self-built & CUDA-Agent & 23 & [12.7,49.2] \\
 & KernelBench & 25 & [20.3,54.1] \\
 & KernelGYM & 25 & [20.3,54.1] \\
 & robust-kbench & 11 & [18.2,38.2] \\
 & DeepReinforce & 25 & [19.2,53.4] \\
 & MusaCoder & 25 & [16.9,52.1] \\
TritonBench-G upstream & all six & 0 & [12.5,12.5] \\
\bottomrule
\end{tabular}
\end{table}

\begin{table}[htbp]
\caption{Pooled abstention bounds (upper panel) and control-axis diagnostics
(lower panel). Delegation and handwritten corpora have zero accepted
\inconcl{} records, so their bounds collapse to the decided rates in
\mainpapertableref{tab:main-fa}{6}. Lower-panel rates use that table's decided
acceptance denominators and describe selected-axis evidence, not a
certificate decomposition.}
\label{tab:appendix-06}
\footnotesize
\setlength{\tabcolsep}{5pt}
\begin{tabular}{@{}llrr@{}}
\toprule
Pool & Verifier & Accepted \inconcl{} & FA bounds\,(\%) \\
\midrule
Main pooled & CUDA-Agent & 30 & [49.2, 61.5] \\
 & KernelBench & 47 & [60.1, 74.0] \\
 & KernelGYM & 59 & [60.1, 75.3] \\
 & robust-kbench & 43 & [60.3, 73.8] \\
 & DeepReinforce & 38 & [56.7, 69.2] \\
 & MusaCoder & 23 & [53.9, 63.0] \\
Natural-5 pooled & CUDA-Agent & 55 & [41.8, 57.6] \\
 & KernelBench & 74 & [52.2, 68.3] \\
 & KernelGYM & 86 & [53.1, 69.9] \\
 & robust-kbench & 55 & [53.2, 66.3] \\
 & DeepReinforce & 65 & [49.3, 64.5] \\
 & MusaCoder & 49 & [45.4, 58.7] \\
\bottomrule
\end{tabular}
\par\smallskip
\begin{tabular}{@{}llrr@{}}
\toprule
Corpus & Verifier & Sem.-evidence\,\% & Impl.-evidence\,\% \\
\midrule
Delegation controls & CUDA-Agent & 27.1 & 86.4 \\
 & KernelBench & 26.7 & 86.7 \\
 & KernelGYM & 26.7 & 86.7 \\
 & robust-kbench & 18.6 & 81.4 \\
 & DeepReinforce & 26.7 & 86.7 \\
 & MusaCoder & 0.0 & 33.3 \\
robust-kbench handwritten & all six & 0.0 & 28.6 \\
\bottomrule
\end{tabular}
\end{table}

Per-corpus axis rates are in the released \texttt{false\_accepts} cells under
\texttt{axis\_FA\_rates}. Evidence-source distinctions follow Appendix~\ref{s:inconcl}.

\paragraph{Nominal-Input Diagnostics and Repeated Execution.}
\label{s:nominal-diagnostics}
The diagnostic set is the record-level union of selected semantic FAILs
accepted by at least one weak verifier; each record is counted once.
Its 62 records all have FAIL certificates: 16 each from CUDA-L1,
AI-CUDA-Engineer, and the delegation controls, 10 from TritonBench-G's
self-built layer, and 4 from KernelBenchX. IDs appear in
\texttt{nominal\_strict\_comparator} in the released table data.

The input-provider contract determines the executable policy schedule.
For these 62 records, contract adaptation replaces ten unsupported optional
transformation slots in the 12-plus-4 base schedule with native-seed slots
before execution. Each instantiated plan contains \texttt{value\_normal},
14 \texttt{native\_seed\_*} policies, and one \texttt{repeated\_run} policy
with three calls. The nominal diagnostic excludes \texttt{repeated\_run}
and examines the other 15 stored policy results.
The nominal signal rule and aggregate counts are reported in
\mainpapersectionref{subsec:parametric-structural}{5}.

The four non-numeric signals are NaN/Inf against a finite reference
(\texttt{sakana\_\_L2\_T80}); fp16 output where fp32 is expected
(\texttt{cuda\_l1\_\_L1\_T8}, \texttt{cuda\_l1\_\_L2\_T96}); and a
return-structure mismatch, expected \texttt{tuple[2]} versus the
\texttt{linalg\_eig} return type (\texttt{kernelbenchx\_\_eig}).
The latter is a structural diagnostic signal; whether it constitutes a
violation depends on the declared output contract.

For example, \texttt{kernelbenchx\_\_tanh} has maximum relative error
0.3393 but maximum absolute error $1.7881\times10^{-7}$; it passes with
$\rho_\ell=10^{-5}$ and $a_\ell=2.4424\times10^{-7}$.

The record without either nominal signal, \texttt{kernelbenchx\_\_exp\_sqrt},
has repeated-run evidence in the complete record selected for both its
certificate and semantic axis. Its first invocation passes (maximum relative
error $1.6746\times10^{-7}$); the second and third fail (17.8396 each),
with maximum absolute error 3.46937 and 75\% of elements beyond tolerance.

\section{\inconcl{} Evidence Categories}
\label{s:inconcl}

Table~\ref{tab:s3-root-causes} breaks down the unresolved-axis evidence
summarized in \mainpapersectionref{subsec:inconclusive}{6.5}.

\begin{table}[htbp]
\caption{Evidence labels for 445 unresolved record--axis entries across 286
records. Multiple axes can contribute labels within a record; these labels
do not assign responsibility.}
\label{tab:s3-root-causes}
\scriptsize
\setlength{\tabcolsep}{4pt}
\begin{tabular}{@{}p{0.58\linewidth}rp{0.26\linewidth}@{}}
\toprule
Axis-level evidence label & Count & Interpretation \\
\midrule
Reference-calibration ceiling & 132 & reference/evidence boundary \\
Unresolved after verified execution & 90 & execution evidence incomplete \\
Execution timeout & 51 & execution evidence incomplete \\
Candidate artifact missing or syntax error & 51 & candidate artifact boundary \\
Parameter/state mapping unproven & 49 & semantic precondition \\
Candidate build diagnostic & 26 & candidate artifact boundary \\
Reference precision or output unsupported & 20 & reference boundary \\
GPU memory shortage & 18 & execution environment boundary \\
Reference nonfinite output & 6 & reference boundary \\
Insufficient work/task evidence & 2 & reference trace lacks compute \\
\bottomrule
\end{tabular}
\end{table}

\paragraph{State-Alignment Evidence.}
An example behind the state-correspondence requirement in
\mainpapersectionref{subsec:semantic-axis}{3.2} is replacement of reference
modules by directly held \texttt{nn.Parameter} objects.
The 49 ``parameter/state mapping unproven'' labels comprise 17 semantic,
16 implementation, and 16 measurement entries across 18 records; they
count unresolved evidence entries, not separate candidate defects.

Certificate and diagnostic-axis conclusions retain separate evidence sources.
In the ledger, 140 FAIL certificates coexist with 158 unresolved selected
axes. Four FAIL certificates have no selected FAIL axis, and one \inconcl{}
certificate has three selected PASS axes. These five certificates and their
diagnostic axes cite different source records, so the displayed axes cannot
reconstruct the certificate or explain its abstention.
Category evidence may also cite a source different from the selected axis;
the released per-axis ledger and evidence strings preserve this distinction.
The two insufficient-work labels concern absent reference compute traces,
not the total number of R-A/R-B exclusions.

For \texttt{cuda\_l1\_\_L2\_T83}, execution \texttt{1cb62964\ldots}
supplies three PASS axes. R-B changes its stored PASS certificate to effective
\inconcl{} without changing those axes. During source export, the later input
\texttt{2e358987\ldots} is an all-\inconcl{} worker placeholder reporting a
\texttt{SchemaError} for a missing candidate-state hash. The $D/U$ rule retains
the execution's diagnostic sources; the $U/U$ rule binds the certificate to the
placeholder. R-B explains ineligibility, while input-order replacement explains
the selected source. The placeholder's error neither overturns the PASS axes
nor supplies evidence for R-B or a candidate defect.

\section{Measurement Evidence and RQ3 Timing Details}
\label{s:measurement}

The following tables distinguish fixed-source timing guards and
source-pattern observations from the selected-axis summary in
\mainpapersectionref{subsec:measurement-axis-results}{6.2}.

\begin{table}[htbp]
\caption{Measurement observations and static-source review.
Timing rows use the fixed 577-record guard cohort within the 716-record audit;
static rows summarize a separate fixed scan and its PASS review subset.
The stream row reports only the ratio component of the criterion in
\mainpapersectionref{subsec:meas-tier}{3.4}.}
\label{tab:appendix-08}
\scriptsize
\setlength{\tabcolsep}{4pt}
\begin{tabular}{@{}p{0.60\linewidth}>{\raggedleft\arraybackslash}p{0.33\linewidth}@{}}
\toprule
Item & Value \\
\midrule
Hidden-stream ratio in the fixed guard cohort: p50 / p95 & 1.09 / 1.26 \\
Thread-injection / lazy-evaluation signals & 0 / 0 \\
Static source-pattern matches in fixed scan (non-gating) & 122 records \\
Flagged PASS subset selected for source review & 19 records \\
Source-text categories in reviewed subset & 16 dtype-table/handling text,
1 cuDNN TF32 setting, 2 bf16 comments \\
\bottomrule
\end{tabular}
\end{table}

Static source-pattern matches identify review targets without establishing
executed precision. The fixed scan flags 122 records. Pattern counts are
record-level and non-exclusive: fp16 85, bf16 72, two TF32 patterns 17 each,
autocast 14, and \texttt{.half()} 6. One \texttt{cudnn.allow\_tf32=True}
setting can match both TF32 patterns.

Source review selects 19 flagged records marked PASS in the scan input;
each also has a PASS certificate in the evaluation ledger. This scan-defined
subset does not establish the complete set of static matches among the 172
PASS certificates.
Sixteen contain dtype-table or handling text, one a cuDNN TF32 setting,
and two have bf16 matches only in comments. These records contain no
runtime precision evidence, and numerical agreement does not identify the
executed precision mode. Static matches remain non-gating; the constructed
dtype-downgrade probe is detected by the semantic axis.

\paragraph{Guard Coverage and Evidence Sources.}
Fixed-source guard membership and selected measurement verdicts describe
different evidence views of the same 716 records
(Table~\ref{tab:s5-guard-coverage}). Their 577 guard records and 651 PASS
records overlap on 558 IDs. The fixed source omits guards on 139 records:
98 with unresolved state mapping, 40 worker failures, and one timeout.

\begin{table}[htbp]
\caption{Fixed timing-guard membership and selected measurement verdicts
over the same 716 record IDs.}
\label{tab:s5-guard-coverage}
\footnotesize
\begin{tabular}{@{}lrrr@{}}
\toprule
Fixed-source guard membership & Selected PASS & Selected \inconcl{} & Total \\
\midrule
Present & 558 & 19 & 577 \\
Absent & 93 & 46 & 139 \\
\midrule
Total & 651 & 65 & 716 \\
\bottomrule
\end{tabular}
\end{table}

Every selected measurement PASS has passed timing, materialization,
thread, and stream guards, including the 93 records outside the fixed guard
cohort. The 44 records without guard observations in
\mainpapersectionref{subsec:measurement-axis-results}{6.2} comprise 17 state-mapping problems, 25 worker
failures, one audit-context failure, and one timeout. Guard presence therefore
does not itself imply a decided axis. The fixed-source stream quantiles and
mutation census describe the 577-record cohort, not the selected-source
guard cohort; RQ3 uses the former timing population.

\paragraph{RQ3 Timing Exclusions and Corpus Breakdown.}
The 394 withheld speedups reported in \mainpapersectionref{sec:rq3}{7}
comprise 285 exclusions for reference self-variation, 82 for intervals
excluding one, and 27 for other gate conditions.
Table~\ref{tab:rq3-all-corpora} gives the per-corpus certificate breakdown;
all 32 PASS certificates share their complete source record with the
corresponding timing evidence.

\begin{table}[htbp]
\caption{Certificate outcomes among all 180 measurable records.
Measurable denotes A/A passage and a positive recorded local speedup.
Timing records and audit certificates are joined by record ID.}
\label{tab:rq3-all-corpora}
\footnotesize
\begin{tabular}{@{}lrrrr@{}}
\toprule
Corpus & Measurable & PASS & FAIL & \inconcl{} \\
\midrule
CUDA-L1 & 64 & 0 & 54 & 10 \\
AI-CUDA-Engineer & 68 & 22 & 39 & 7 \\
KernelBenchX & 2 & 0 & 2 & 0 \\
TritonBench-G (self-built ref.) & 22 & 8 & 4 & 10 \\
TritonBench-G (upstream ref.) & 1 & 0 & 1 & 0 \\
Delegation controls & 20 & 1 & 19 & 0 \\
robust-kbench handwritten & 3 & 1 & 2 & 0 \\
\midrule
Total & 180 & 32 & 121 & 27 \\
\bottomrule
\end{tabular}
\end{table}

Paired summaries weight records equally without assuming independent tasks.
For the 132-record main paired cohort, the median of record-wise
measured/claimed ratios is 0.934838, whereas the ratio of marginal medians
is 0.852335.

\paragraph{RQ3 Certificate Sources and Attribution.}
Table~\ref{tab:rq3-source-linkage} expands the 54 source-mismatched main-cohort
records identified in \mainpapersectionref{sec:rq3}{7}.
Ten diagnostic records (all \inconcl{}) also have different certificate and
timing hashes; all eight diagnostic PASS records share theirs.
A different complete-record hash does not itself imply different candidate
source code. Policy fingerprints match across the 54 main-cohort links,
while audit-program, configuration, and environment fingerprints differ.
Matching recorded identity fields therefore does not establish identical
execution contexts or realized execution paths.

\begin{table}[tb]
\caption{Source correspondence for the 54 main-cohort records with
different timing and certificate record hashes. All are uncertified.
}
\label{tab:rq3-source-linkage}
\footnotesize
\setlength{\tabcolsep}{3pt}
\begin{tabular}{@{}r p{0.21\linewidth} p{0.70\linewidth}@{}}
\toprule
$n$ & Records & Confirmed correspondence and limits \\
\midrule
45 & Other linked records & Recorded candidate, task, contract, and reference
fingerprints, plus canonical input and pre-/post-state hashes, match. \\
1 & CUDA-L1:\newline\texttt{L3\_T46} & Same candidate source; the candidate-fingerprint
difference is explained by four batch-normalization state-adaptation metadata entries. \\
3 & AI-CUDA-Engineer:\newline\texttt{L3\_T37}, \texttt{L3\_T39}, \texttt{L3\_T40} &
Candidate, task, and input match. Recorded state coverage differs in the
ordinary attributes \texttt{h0/c0} for T37 and \texttt{h0} for T39/T40. This correspondence does not
establish full state equality or a change of reference code. \\
1 & CUDA-L1:\newline\texttt{L2\_T80} & Candidate, task, reference, and input hashes
match; the certificate source lacks candidate canonical pre-/post-state hashes. \\
4 & CUDA-L1:\newline\texttt{L1\_T48}, \texttt{L1\_T51}, \texttt{L1\_T52}, \texttt{L2\_T83} &
Timeout or validation-failure certificate records retain linked candidate code
but lack complete input/state execution identity. \\
\bottomrule
\end{tabular}
\end{table}

\paragraph{RQ3 Duration Strata.}
A/A passage varies with reference duration in the fixed 577-record cohort.
Median reference durations are
2.544208 ms for the 183 A/A passes and 0.112656 ms for all 394 withholdings.
The 285 records withheld for excessive reference CV have
median 0.086720 ms. In selected duration strata, A/A pass counts are
63/362 (17.40\%) below 0.5 ms,
64/105 (60.95\%) in $[2,10)$ ms, and 34/66 (51.52\%) at 10 ms or above.
Together, the latter bins have 98/171 passes (57.31\%); the bin-specific
rates are observed proportions, not confidence bounds. These timing
associations do not identify the causal contributions of kernel properties,
host conditions, or measurement overhead.

\section{Extended Case Studies}
\label{s:cases}

\paragraph{Launch-Provenance Controls.}
The launch controls in \mainpapertableref{tab:controls}{7} have the following
per-case observations (\texttt{launch\_controls\_results.json}).
Columns count forbidden vendor
(Forb.), candidate (Cand.), and mediated library (Med.) launches.

\begin{center}
\scriptsize
\setlength{\tabcolsep}{3pt}
\begin{tabular}{@{}p{0.24\linewidth}p{0.40\linewidth}rrr@{}}
\toprule
Case & Construction & Forb. & Cand. & Med. \\
\midrule
C-LP1 direct vendor call & extension calls \texttt{cublasSgemm} below the
dispatcher & \textbf{2} & 0 & 0 \\
C-LP2 mediated call & plain \texttt{torch.mm} forward & 0 & 0 &
\textbf{2} \\
C-LP3 self-written & hand-written CUDA matmul kernel & 0 & \textbf{1} &
0 \\
C-LP4 name collision & custom kernel with vendor-style name
(\texttt{ampere\_sgemm\_...\_like\_custom}) & 0 & \textbf{1} & 0 \\
\bottomrule
\end{tabular}
\end{center}

The controls distinguish the tested launch origins.
C-LP1 matches two inventory symbols: an
\texttt{ampere\_sgemm} variant and a \texttt{splitKreduce} kernel. C-LP2
places the tested dispatcher-mediated kernels in the library bucket. This
launch-channel negative remains subject to Channel 1's contract assessment.
C-LP4's vendor-style name produces no false vendor attribution.
Comparisons against \texttt{torch.mm} record maximum absolute differences
of $0$, $0$, $3.8147\times10^{-5}$, and $4.1962\times10^{-5}$ for C-LP1--4,
respectively. These observations have no recorded numerical acceptance
threshold; the 4/4 result tests launch-bucket expectations.

\paragraph{Evidence-Channel Attribution Matrix.}
The matrix allocates 313 selected implementation FAIL records across the
two evidence channels; independent validity is assessed in Appendix~\ref{s:annotation}.
Of the 229 main-corpus records with selected implementation FAIL evidence
identified in \mainpapersectionref{subsec:impl-channels}{6.1}, 23 also carry
selected semantic FAIL evidence. These are record-level
overlaps: selected axes and certificates need not share an execution source.
No main-corpus record receives \cthreea{} under the reporting priority:
semantic failures can coexist with \cthreeb{}, while semantic-only false
accepts without a qualifying signature receive \ctwo{}. These codes do not
establish author intent.

\begin{center}
\scriptsize
\setlength{\tabcolsep}{4pt}
\begin{tabular}{@{}lrrrr@{}}
\toprule
Corpus & Impl.\ FAIL & Dispatcher only & Launch only & Both \\
\midrule
CUDA-L1 & 136 & 127 & 1 & 8 \\
AI-CUDA-Engineer & 93 & 89 & \textbf{4} & 0 \\
KernelBenchX & 22 & 22 & 0 & 0 \\
TritonBench-G (both layers) & 8 & 8 & 0 & 0 \\
Delegation controls & 52 & 52 & 0 & 0 \\
robust-kbench controls & 2 & 1 & \textbf{1} & 0 \\
\midrule
Total & 313 & 299 & 6 & 8 \\
\bottomrule
\end{tabular}
\end{center}

\paragraph{Case 1 (Extended): \texttt{llama\_ffw}, a Policy-Sensitive
Diagnostic.}
For the \texttt{llama\_ffw} case in
\mainpapersectionref{subsec:impl-channels}{6.1}, the reference matmul
signature is $1.8{\times}10^{11}$ FLOPs.
Channel 2 observes five device activities: three vendor-kernel launches,
one custom-kernel launch, and one memset. The vendor SGEMM kernels exactly match
the sealed inventory: two \texttt{ampere\_sgemm\_128x32\_tn} and one
\texttt{ampere\_sgemm\_128x64\_tn}. Saved CUDA source contains three
\texttt{cublasSgemm} calls, consistent with the runtime hits. The saved
wrapper matches the record's wrapper fingerprint; this match does not fully
identify the bytes of the external CUDA source used at execution.
GPU-time buckets attribute 5{,}435\,$\mu$s to vendor work
and 21\,$\mu$s (0.4\%) to the candidate's
\texttt{fused\_silu\_mul\_\allowbreak predicate\_kernel}: the custom work is the SiLU
epilogue, while cuBLAS performs the core GEMMs.
The recorded implementation signature is
\texttt{launch:\allowbreak forbidden\_library:3}.

\paragraph{Case 2 (Extended): Silent-Fallback Shell
\texttt{cuda\_l1\_\_L2\_T46}.}
After the custom-convolution build fails in the frozen environment, the
candidate's wrapper executes native PyTorch fallback.
Channel 1 records dominant \texttt{aten::convolution} work
(estimated $1.50{\times}10^{12}$ FLOPs; reference $5.99{\times}10^{11}$).
Joint noise replacement of convolution outputs returns LEAK: maximum relative output
change $1.84{\times}10^{7}$, leakage fraction 1.0. Channel 2 records three
non-mediated vendor launches: two cuDNN layout transforms
(\texttt{cudnn::\allowbreak ops::\allowbreak nchwToNhwcKernel}) and one implicit-GEMM launch.
KernelBench, KernelGYM, and robust-kbench accept; CUDA-Agent rejects for
runtime \texttt{F.*} blocking, MusaCoder for compilation failure, and
DeepReinforce for output mismatch at $10^{-2}$ tolerance, with underlying
cause undetermined.

\paragraph{Case 3 (Extended): Attribution.}
In the TritonBench-G upstream layer,
\texttt{124\_logsumexp\_fwd} is convicted only after \texttt{do(P:=noise)}
confirms output dependence on \texttt{aten::logsumexp} (all six weak
verifiers accept it).

The implementation-axis misses summarized in
\mainpapertableref{tab:controls}{7} comprise three GEMV-level probes below
the targeting threshold and five transposed-convolution probes on
unknown-operation paths. For the GEMV probes, reference-to-candidate
work-proxy ratios are 192, 323, and 384; these compare estimator outputs,
not measured hardware FLOPs.
Four transposed-convolution probes are captured by the semantic axis;
one (\texttt{convtrans2d\_k2s2\_exact}) passes all axes.

\section{Oracle-Validity Details}
\label{s:validity}

\paragraph{Constructed Controls, per Axis.}
Beyond the family-level results in \mainpapertableref{tab:controls}{7},
the stale-cache and shape-specialization controls receive semantic PASS
but implementation and certificate FAIL. Each executes \texttt{value\_normal},
14 native-seed policies, and \texttt{repeated\_run} with three calls.
This 16-slot plan does not establish alternative-shape coverage;
the causes of these two semantic misses remain undetermined.

Measurement controls separate decisive violations from non-gating signals.
Of six positives, two timing-tamper probes yield measurement FAIL, three
thread/lazy/precision probes raise their designed non-gating signals with
measurement PASS, and one hidden-stream probe produces no signal.
Thus 5/6 activate a signal, while 2/6 establish measurement violations;
both measurement negatives have PASS with no guard signal.

\paragraph{Upstream Reference Quality.}
For the upstream-reference probe in
\mainpapertableref{tab:oracle-landscape}{5}, a FAIL certificate alone does
not attribute a defect to the upstream reference or its author.
Such attribution requires record-level evidence separating candidate,
reference, and task/input causes; execution failure alone is insufficient.

\section{RQ4 Details}
\label{s:rq4}

\paragraph{Registration and Static-Screening Details.}
The task pool and slot order are specified in
\mainpapersectionref{sec:selector-intervention}{8}.
The CUDA reference-basename join uses level lists of 46, 140, and 20 tasks;
14 joined tasks lie outside the resource plan.
The same basename join is used for the TritonBench G/pref intersection;
this join key does not establish semantic equivalence of registered references.
Historical static screening flagged 191 slots for simultaneous explicit
\texttt{PYBIND11\_MODULE} and \texttt{load\_inline(functions=...)} bindings,
which can duplicate generated C++ exports. This compilation issue is distinct
from tensor-storage aliasing; membership in a skip list does not prove that
no earlier execution was attempted.

\paragraph{Reconciled Evidence and Replay Eligibility.}
The 204 reused slots described in
\mainpapersectionref{sec:selector-intervention}{8} come from 181 distinct
complete primary-audit records: 23 TritonBench records each supply two slots.

Let $E_W$, $E_{\mathrm{axis}}$, and $E_1$ denote the weak, axis-only, and
reconciled eligible-slot sets defined by the rules in
\mainpapersectionref{sec:selector-intervention}{8}.
There are 186 axis-PASS slots and 184 eligible slots. The two remaining
axis-PASS exclusions are historical Opus c0 and c2 on
\texttt{cuda\_l1\_\_L2\_T80}. This contrast does not isolate R-A's effect.

\paragraph{Timing-Cache Eligibility and Provenance.}
Beyond the measurement and guard conditions in
\mainpapersectionref{sec:selector-intervention}{8}, timing requires successful
state alignment and available task tolerances and calibration settings.
A/A requires \texttt{accepted=True} under the full protocol in Appendix~\ref{s:params}.
Local A/B timing uses ABBA pairing and requires timing PASS together with
successful output comparison, cross-checkpoint consistency, state/input
checks, and stability checks. Warmup counts, batch sizes/counts, and stability
limits match the A/A settings in \mainpapersectionref{subsec:meas-tier}{3.4}.

A usable cache entry has \texttt{PerformanceFacts.status=PASS} and positive,
finite \texttt{speedup}, \texttt{ref\_ms}, and \texttt{cand\_ms}; otherwise
its public speedup is \texttt{null}. All 68 non-null speedups satisfy these
conditions and are computed as
\[
 \mathrm{speedup}=\frac{\operatorname{median}(\text{local reference times})}
 {\operatorname{median}(\text{local candidate times})}.
\]
Ranking reads non-null cached speedups without rerunning these checks;
upstream speedup claims are excluded.

Timing availability and shipping eligibility impose different conditions.
Measurement PASS alone does not provide a usable speedup: any activated
measurement guard, including the 114 stream signals, withholds it.
Conversely, timing availability does not require semantic or implementation
PASS; shipping eligibility uses $E_W$, $E_{\mathrm{axis}}$, or $E_1$.

\paragraph{Ordering Details.}
Each gate selects from the fixed pool rather than post-filtering $S_0$'s choice.
Exact \texttt{PerformanceFacts.\allowbreak speedup} ties use the earlier slot,
without tolerance or a secondary key.
Selection neither minimizes raw candidate latency nor requires speedup above one.

\begin{table}[htbp]
\caption{Outcome distribution over 1{,}160 slots. The R-C attribution results
are in \mainpapersectionref{sec:selector-intervention}{8}.
Convicted slots rejected by all six weak verifiers are counted separately;
otherwise the reported code is shared by the accepting verifiers.}
\label{tab:appendix-14}
\scriptsize
\setlength{\tabcolsep}{4pt}
\begin{tabular}{@{}lrrp{0.42\linewidth}@{}}
\toprule
Outcome & Slots & Share & Note \\
\midrule
Eligible under $S_1$ & 184 & 15.9\% & covers 62 of 232 tasks \\
FAIL; all six reject & 48 & 4.1\% & aggregate rejection count \\
\ctwo{} (silent semantic defect) & 17 & 1.5\% & \\
\cthreea{} (semantic violation w/ signature) & 0 & 0.0\% & \\
\cthreeb{} (implementation violation) & 122 & 10.5\% & includes reused primary evidence \\
\cthreec{} (measurement violation) & 0 & 0.0\% & \\
\inconcl{} & 789 & 68.0\% & unavailable evidence or withheld eligibility \\
\bottomrule
\end{tabular}
\end{table}

\paragraph{Shipping Estimands.}
The slot totals do not determine how much task-level abstention is attributable
to candidate violations or unavailable evidence.
Shipping rates distinguish tasks from deliveries. For
$S\in\{S_0,S_{\mathrm{axis}},S_1\}$, let $d_t(S)$ indicate delivery on
task $t$, and let
$h_t(S)$ indicate a human-labeled unsafe shipment, with $h_t(S)=0$
when $S$ abstains. On this 232-task pool, coverage is
$C(S)=\sum_t d_t(S)/232$, and the human unsafe-task rate is
$U(S)=\sum_t h_t(S)/232$. For nonzero coverage,
$V(S)=\sum_t h_t(S)/\sum_t d_t(S)$ is the unsafe fraction among deliveries,
and $U(S)=C(S)V(S)$. For an unknown shipment label, the lower and upper
bounds assign $h_t(S)=0$ or $1$, respectively; these are identification bounds,
not confidence intervals. A smaller unsafe-task rate can reflect fewer deliveries as well as a
different mix of delivered candidates.
Oracle-visible unsafe rates replace the human indicator with the
corresponding audit-label indicator; the two standards are not interchangeable.

\paragraph{Gate and Ordering Contrasts.}
For each weak verifier, let $u_{w,f}$, $u_{w,v}$, $u_{c,f}$,
and $u_{c,v}$ be oracle-visible unsafe shipments divided by 232 for
$S_0$, $S_0$-fast, $S_1$-first, and $S_1$, respectively.
Here $w/c$ denote weak/audit gates and $f/v$ denote first/fastest
ordering. Unresolved shipments and abstentions remain in the denominator
without entering the unsafe numerator; neither is thereby labeled safe.
The two descriptive reductions, in percentage points, are
\[
G=\frac{100}{2}\bigl[(u_{w,f}-u_{c,f})+(u_{w,v}-u_{c,v})\bigr],
\qquad
R=\frac{100}{2}\bigl[(u_{w,f}-u_{w,v})+(u_{c,f}-u_{c,v})\bigr].
\]
We average $G$ and $R$ equally over the six verifiers, obtaining
30.927 and 0.826 pp. Per-verifier $G$ ranges from 10.345 to 47.845 pp.
These six contrasts share tasks and are not independent samples.
Both audit-gated cells have zero oracle-visible unsafe shipments
by their eligibility definition, so $R$ is half the weak-gate ordering
contrast.

\paragraph{Delivery with Speedup.}
Table~\ref{tab:s8-selector-cells} gives the per-verifier SafeFast breakdown
behind \mainpapersectionref{sec:selector-intervention}{8}.
Missing speedup evidence contributes zero to this endpoint without
establishing absence of a speed advantage. The small ordering contrast on
unsafe counts therefore does not imply absence of benefit on speed-qualified
delivery.

\begin{table}[htbp]
\caption{SafeFast counts by verifier and ordering on the same 232 tasks.
Pairs give counts at $k{=}1,2$ and require membership in $E_1$ and credible
speedup $>k\times$. Corresponding oracle-visible unsafe counts are in
\mainpapertableref{tab:selector-results}{11}.}
\label{tab:s8-selector-cells}
\footnotesize
\setlength{\tabcolsep}{6pt}
\begin{tabular}{@{}lrr@{}}
\toprule
Weak verifier & $S_0$ SafeFast (1, 2) & $S_0$-fast SafeFast (1, 2) \\
\midrule
CUDA-Agent & (0, 0) & (3, 2) \\
KernelBench & (0, 0) & (3, 2) \\
KernelGYM & (0, 0) & (3, 2) \\
robust-kbench & (0, 0) & (3, 2) \\
DeepReinforce & (0, 0) & (3, 2) \\
MusaCoder & (3, 1) & (5, 2) \\
\midrule
Shared audit gate & $S_1$-first SafeFast (1, 2) & $S_1$ SafeFast (1, 2) \\
All six verifiers & (4, 2) & (6, 2) \\
\bottomrule
\end{tabular}
\end{table}

\begin{table}[htbp]
\caption{Runtime diagnostics. Worker durations and cache-write intervals
have different units of observation; neither is a direct deployment-cost
measurement.}
\label{tab:appendix-16}
\scriptsize
\setlength{\tabcolsep}{4pt}
\begin{tabular}{@{}p{0.34\linewidth}p{0.41\linewidth}r@{}}
\toprule
Observation & Sample & Statistic \\
\midrule
Worker duration with end record & Primary audit: 729 attempts, including repeated attempts &
p50/p95: 1.025/18.081 min \\
Strong-cache write interval & RQ4 46-task segment: 196 intervals;
endpoints on 44 tasks, 196 candidates & p50: 7.460 min \\
Weak-cache write interval & Same segment: 887 intervals;
endpoints on 45 tasks, 174 candidates; six verifiers & p50: 77.142 s \\
Cached first-PASS scans & 72 tasks, 360 slots; six weak verifiers &
wall time/slot: 0.03 ms \\
\bottomrule
\end{tabular}
\end{table}

\paragraph{Cost Measurement and Scope.}
Runtime summaries retain their observation units and sampling filters.
Worker durations include each attempt with an end record, including repeats;
their quantiles are not per unique kernel. Cache intervals are differences
between adjacent writes, retaining $0<\Delta\leq3600$\,s for the strong
cache and $0<\Delta\leq300$\,s for the weak cache. For the reported
quantiles $q\in\{0.5,0.95\}$, the script selects zero-based index
$\lfloor nq\rfloor$ among $n$ sorted observations.
The p50 ratio is $7.460\times60/77.142\approx5.80$.
Different filters and candidate coverage make these unpaired summaries;
intervals do not isolate an individual audit stage's start and end.

Using the strong-cache p50 for one audit per task is a cost proxy.
Beyond the cost limitations in \mainpapersectionref{sec:selector-intervention}{8},
per-axis audit costs and human reference-authoring costs are unmeasured.

The cached-scan measurement covers data loading, indexing, and first-PASS
selection. Its 0.03 ms per slot is total wall time divided by 360 over
72 tasks and six weak verifiers, including reads of 12 audit JSON files,
index construction, and scans for the first weak-PASS and first stored
certificate-PASS entries. It excludes evidence collection and does not
measure fastest ordering or the refreshed $S_1$ workflow.

\section{Blind-Annotation Details}
\label{s:annotation}

The historical blind review comprises 450 independently annotated units:
120 validity-sample records and 330 shipment-census candidates. The refreshed
RQ4 union adds one unannotated candidate; it is not included in the following
annotation-reliability statistics. Packets randomize identifiers and omit
oracle axes/certificates, weak verdicts, timing ratios, and source-provenance
fields. Annotators inspect candidate/reference artifacts and independent
executions for correctness, delegation, and measurement.
The 102 units with at least one dimension disagreement (29 validity units
and 73 census units, concentrated in delegation declarations) receive
third-party arbitration with a written rationale for each unit.
A unit may disagree on multiple dimensions: the 102 units yield 119
dimension-level rulings, adopting annotator A 49 times and B 70 times.
All rulings use one of the two annotators' labels.

\paragraph{Policy and Human-Reference Scope.}
Upstream prompts request CUDA optimization: KernelBench (Appendix C.1 of the upstream paper) permits partial
operator replacement, and the examined CUDA-L1 (Table 3 of the upstream paper) and AI-CUDA-Engineer (Appendix B.2 of the upstream paper) prompts
do not state a universal ban on core library computation. The frozen study policy imposes core
self-implementation; e.g., \texttt{cuda\_l1\_\_L1\_T1} and
\texttt{sakana\_\_L1\_T1} have \texttt{requires\_self\_impl=true}.
The policy/rubric distinction is discussed in
\mainpapersectionref{sec:experimental-design}{4} and
\mainpapersectionref{subsec:oracle-validity}{6.3}.
On weak-accepted, oracle-decided main-corpus records, human-referenced
false-accept estimates are 8.7--23.6 percentage points below corresponding
oracle rates; these differences do not identify their causes.

\paragraph{Reliability.}
Raw agreement and Cohen's $\kappa$ jointly describe annotation reliability.
Agreement/$\kappa$ by dimension are: validity sample
(correctness 90.8\%/0.740, delegation 80.0\%/0.582, measurement
96.7\%/$-0.006$); shipment census (correctness 95.2\%/0.568, delegation
81.2\%/0.549, measurement 99.4\%/0.497). Measurement labels are highly imbalanced: almost all are honest, with
4 and 2 disagreements, respectively. Because $\kappa$ adjusts for chance
agreement using the annotators' marginal label frequencies, high raw
agreement can coexist with low $\kappa$; neither measure is interpreted alone.

\begin{table}[htbp]
\caption{Oracle-validity sample design. Rows are the frozen original
certificate $\times$ corpus sampling strata; selected certificates define
metric cells. The table reports population $N_h$,
draws $n_h$, human labels defect/clean/unknown, and inclusion probability
$\pi_h$.}
\label{tab:appendix-17}
\scriptsize
\setlength{\tabcolsep}{4pt}
\begin{tabular}{@{}llrrrr@{}}
\toprule
Design certificate & Corpus & $N_h$ & $n_h$ & d/c/u & $\pi_h$ \\
\midrule
FAIL & AI-CUDA-Engineer & 107 & 14 & 12/2/0 & 0.131 \\
FAIL & CUDA-L1 & 79 & 11 & 8/3/0 & 0.139 \\
FAIL & cuDNN-real & 56 & 8 & 8/0/0 & 0.143 \\
FAIL & KernelBenchX & 37 & 5 & 3/2/0 & 0.135 \\
FAIL & robust-kbench & 3 & 2 & 1/1/0 & 0.667 \\
FAIL & TritonBench-G self-built & 21 & 3 & 1/2/0 & 0.143 \\
FAIL & TritonBench-G upstream & 15 & 2 & 2/0/0 & 0.133 \\
INCONCL & AI-CUDA-Engineer & 31 & 3 & 2/1/0 & 0.097 \\
INCONCL & CUDA-L1 & 162 & 19 & 11/8/0 & 0.117 \\
INCONCL & KernelBenchX & 13 & 2 & 0/1/1 & 0.154 \\
INCONCL & TritonBench-G self-built & 39 & 4 & 0/2/2 & 0.103 \\
INCONCL & TritonBench-G upstream & 4 & 2 & 1/0/1 & 0.500 \\
PASS & AI-CUDA-Engineer & 91 & 26 & 0/26/0 & 0.286 \\
PASS & cuDNN-real & 4 & 2 & 2/0/0 & 0.500 \\
PASS & KernelBenchX & 13 & 4 & 0/4/0 & 0.308 \\
PASS & robust-kbench & 5 & 2 & 0/2/0 & 0.400 \\
PASS & TritonBench-G self-built & 29 & 9 & 0/9/0 & 0.310 \\
PASS & TritonBench-G upstream & 7 & 2 & 0/2/0 & 0.286 \\
\bottomrule
\end{tabular}
\end{table}

\paragraph{Validity-Sample Confusion (Unweighted, Certificate Level).}
The unweighted cells are PASS (44 clean, 3 defect), FAIL (40 defect,
13 clean, 1 unknown), and INCONCL (8 defect, 8 clean, 3 unknown).
Human defect is reference-positive and certificate FAIL is audit-positive;
the weighted metric denominators are specified in
\mainpapersectionref{subsec:oracle-validity}{6.3}.

\paragraph{Interval Method.}
Point estimates are ratios of counts weighted by $1/\pi_h$; 95\% model-based
posterior intervals use finite-population stratified Jeffreys--Dirichlet
completion. For validity
metrics, $h$ denotes original certificate $\times$ corpus and $j$ the
selected final certificate cell. Let $x_{hj}$ count observed defect, clean,
and unknown labels, with population size $N_{hj}$ and sample size $n_{hj}$.
For replicate $b$,
\[
\begin{gathered}
x_h=\sum_j x_{hj},\qquad
p_h^{(b)}\sim\mathrm{Dirichlet}(x_h+\tfrac12\mathbf{1}_3),\\
Y_{hj}^{(b)}\mid p_h^{(b)}\sim
\mathrm{Multinomial}(N_{hj}-n_{hj},p_h^{(b)}),\qquad
\widetilde{x}_{hj}^{(b)}=x_{hj}+Y_{hj}^{(b)}.
\end{gathered}
\]
Each stratum shares one $p_h^{(b)}$ across its cells; cell completions are
conditionally independent given that draw. Observed labels, including
unknowns, remain fixed.

The uncertainty scope and unassessed exchangeability sensitivity are stated in
\mainpapersectionref{subsec:oracle-validity}{6.3}.

For human-referenced false-accept rates, $h$ is further crossed with weak
acceptance, using the same shared-draw construction. The decided estimate uses accepted records with
certificate PASS or FAIL; the all-accepted lower and upper bounds assign
unknown labels to clean and defective, respectively. The all-accepted lower
and upper point estimates coincide because sampled accepted units have no
unknown labels; posterior
completion retains unknown-label probability, yielding distinct intervals.
Metrics are recomputed on $B{=}2000$ completed populations. Zero-event
strata with unsampled units retain residual uncertainty. Each corpus contributes at most one
record per task; source kernels may recur across reference layers. The validity completion simulation uses
seed 20260817; the human-referenced false-accept simulation uses seed 20260826.

\paragraph{Statistical Reproducibility.}
The statistical inputs bind each of the 120 validity-sample units' dimension-level
and combined human labels to its original stratum, $N_h$, $n_h$, inclusion
probability, and selected certificate.
Hash and join checks bind the inputs. Calculations
use the retained six-decimal inclusion probabilities; exact $n_h/N_h$
values are also retained. An independent standard-library implementation
reproduces the six validity point estimates and 24 interval sets: six
validity intervals and three human-referenced false-accept intervals per
verifier. This recomputation verifies computation conditional on the stated model.

\begin{table}[htbp]
\caption{Human-referenced false-accept rates over all accepted main-corpus
records. Lower/upper intervals assign unknown labels to clean/defect.
The point estimates coincide under these assignments; corresponding
oracle-decided estimates and intervals are in
\mainpapertableref{tab:main-fa}{6}.}
\label{tab:appendix-18}
\scriptsize
\setlength{\tabcolsep}{5pt}
\begin{tabular}{@{}lrrr@{}}
\toprule
Verifier & All-accepted point & Lower 95\% interval & Upper 95\% interval \\
\midrule
CUDA-Agent & 44.2\% & [32.4,52.9] & [37.7,57.4] \\
KernelBench & 47.0\% & [33.7,54.4] & [37.6,58.3] \\
KernelGYM & 48.0\% & [36.6,56.7] & [40.7,59.3] \\
robust-kbench & 43.7\% & [29.4,51.9] & [33.4,55.9] \\
DeepReinforce & 52.1\% & [38.7,59.7] & [43.9,63.6] \\
MusaCoder & 50.6\% & [36.6,57.1] & [42.1,60.6] \\
\bottomrule
\end{tabular}
\end{table}

\paragraph{Historical Census Provenance.}
The historical 330-unit census was the complete deduplicated delivery union
of $S_0$, $S_0$-fast, $S_{\mathrm{axis}}$-first, and $S_{\mathrm{axis}}$
under the original saved evidence, each applied for all six weak verifiers.
Archived snapshots before annotation contain that pool, ordering, and mapping;
shipment packets also hid selector membership.
The original reference standard labels 220 of these 330 candidates clean
on all three dimensions. Both refreshed audit selectors choose the same
known unsafe and unreviewed candidates identified in
\mainpapersectionref{sec:selector-intervention}{8}; their coverage and
human-unsafe bounds are reported in
\mainpapertableref{tab:selector-results}{11}.

\paragraph{Case 4: Primary-Audit Evidence and the Independent Witness.}
The historical axis-only selection of CUDA-L1 L2 T80 (BR-171) was labeled
correctness defect and measurement dishonest by both annotators.
Its class-level \texttt{\_\_call\_\_} replacement affects new instances,
whose \texttt{\_zero\_tensor} remains \texttt{None}; held-out and stress
executions returned no candidate tensor against a one-output reference.
These observations establish the human labels independently of the
zero-valued reference. In the primary audit, the certificate-bound execution
terminates at R-B before the semantic context is created, with zero effective
policies. A separate automatic record reports 16 passing policies, but its
instance lifecycle is not established and it contains no arity mismatch.
The refreshed RQ4 record therefore remains certificate-INCONCLUSIVE;
BR-171 is no longer the selected axis-only candidate. This historical
example cannot explain the refreshed gate's human outcomes by R-B alone.

\paragraph{Atrex-Based Screen and Reproducibility.}
The screen and threshold are described in
\mainpapersectionref{subsec:atrex-comparison}{6.4}.
The adapter preserves upstream missing-observation returns;
outer execution errors yield unavailable observations.

Byte/hash checks bind the adapter sources to the executed package; offline
recomputation matches 248 saved observations (240 OV120 over two seeds,
eight controls). Eighteen unavailable observations represent nine OV120
records across both seeds. This recomputation validates stored computation, not complete-harness
equivalence or fresh GPU execution.

\paragraph{Interpreting Low Attribution.}
At both seeds (20260805 and 424243), the same 24 of 48 false-positive
records contain one distinct kernel symbol in the saved attribution output;
launch activity is aggregated by symbol. For KernelBenchX softmax (OV-100),
each seed observation records 2.84\,$\mu$s for \texttt{\_softmax\_kernel}
and 7.17\,$\mu$s end-to-end (about 39.6\%). The numerator comes from
profiling and the denominator from an independent eager benchmark under
the same seed, not from timing the same forward execution.
Neither the human standard nor the implementation axis identifies delegation
for this candidate.
One observed symbol does not exclude delegation.

\end{document}